\documentclass[a4paper]{article}
\usepackage{amsmath}
\usepackage{amssymb}
\usepackage{amsfonts}
\usepackage{amsthm}
\usepackage{amsopn}
\usepackage{graphicx}
\usepackage{tikz}
\usepackage{url}
\usepackage{float}
\usepackage{hyphenat}
\usepackage[normalem]{ulem}
\usepackage{comment}
\usepackage[unicode, pdftex]{hyperref}
\hoffset= - 1.0in         

\def\Tr{{\rm Tr}\,}

\usepackage{xcolor} 
\definecolor{linkcolor}{HTML}{000000} 
\definecolor{urlcolor}{HTML}{000000} 
\definecolor{citecolor}{HTML}{6600FF} 
\hypersetup{pdfstartview=FitH,  linkcolor=linkcolor,urlcolor=urlcolor, citecolor = citecolor, colorlinks=true}

\begin{document}
\title{\vspace{0.1cm}{\Large {\bf On the block structure of the quantum $\mathcal{R}$-matrix \\in the three-strand braids}\vspace{.2cm}}
	\author{{\bf L. Bishler$^{a,b}$}\footnote{mila-bishler@mail.ru}, \
		  {\bf An. Morozov$^{b,c}$}\thanks{andrey.morozov@itep.ru}, \
	   	  {\bf A. Sleptsov$^{b,c,d}$}\footnote{sleptsov@itep.ru}, \
	   	  {\bf Sh. Shakirov$^{e}$}\thanks{shakirov@fas.harvard.edu}}
}
	\date{ }

\maketitle

\vspace{-5.5cm}
\begin{center}
	\hfill ITEP/TH-38/17\\
    \hfill IITP/TH-23/17\\
\end{center}
\vspace{4.2cm}

\begin{center}
	
	$^a$ {\small {\it Moscow State University, Moscow 119991, Russia}}\\
	$^b$ {\small {\it ITEP, Moscow 117218, Russia}}\\
	$^c$ {\small {\it Institute for Information Transmission Problems,
			Moscow 127994, Russia}}\\
	$^d$ {\small {\it Laboratory of Quantum Topology, Chelyabinsk State	University, Chelyabinsk 454001, Russia}}, \\
		
	$^e$ {\small {\it 	Society of Fellows, Harvard University, Cambridge, MA 20138, USA}}\\

\end{center}

\vspace{0.5cm}

\begin{abstract}
	Quantum $\mathcal{R}$-matrices are the building blocks for the colored HOMFLY polynomials. In the case of three-strand braids with an identical finite-dimensional irreducible representation $T$ of $SU_q(N)$ associated with each strand one needs two matrices: $\mathcal{R}_1$ and $\mathcal{R}_2$. They are related by the Racah matrices $\mathcal{R}_2 = \mathcal{U} \mathcal{R}_1 \mathcal{U}^{\dagger}$. Since we can always choose the basis so that $\mathcal{R}_1$ is diagonal, the problem is reduced to evaluation of $\mathcal{R}_2$-matrices. This paper is one more step on the road to simplification of such calculations. We found out and proved for some cases that $\mathcal{R}_2$-matrices could be transformed into a block-diagonal ones. The essential condition is that there is a pair of accidentally coinciding eigenvalues among eigenvalues of $\mathcal{R}_1$-matrix. The angle of the rotation in the sectors corresponding to accidentally coinciding eigenvalues from  the basis defined by the Racah matrix to the basis in which $\mathcal{R}_2$ is block-diagonal is $\pm \frac{\pi}{4}$.
	
\end{abstract}

\vspace{0.3cm}
\section{Introduction}
HOMFLY polynomials (Wilson loop averages in Chern-Simons theory \cite{wit}), especially colored HOMFLY polynomials \cite{clp1, clp2}, receive now a lot of attention. In part it is due to the connections to other theories such as \cite{other18, other19}, also provided by such alternatives to HOMFLY polynomials as hyper- \cite{hyper36, hyper26, alt8} and super- \cite{super25, super37, evolution0} polynomials.

There are different approaches to evaluation of HOMFLY polynomials. Among them the Reshetikhin-Turaev (RT) group-theoretical method \cite{turaev}, the  approach connected with conformal theories \cite{cft}, useful in calculations of two-bridge and arborescent knots \cite{hidint}-\cite{arbr10}, and the evolution method \cite{evolution0, evolution2, evolution}, which can be very effective for series of knots. For very promising alternative approaches, see \cite{alt8, alt9, alt10}.  In this paper the RT method is used. The polynomial in this method is given by the formula:
\begin{equation}
H^{\mathcal{K}}_{T_1\otimes T_2\ldots}=\mathrm{Tr}_{T_1\otimes T_2\ldots} \prod\limits_{\alpha}\mathcal{R}_{\alpha},
\label{trace}
\end{equation}
where $\alpha$ enumerates all crossings in the braid.
To use the RT approach one should redraw a projection of a knot as a braid. Each line in the braid is associated with a finite-dimensional irreducible  representation $T_i$ of the quantum group $SU_q(N)$. In our paper we worked with three-strand braids. Since we want to calculate polynomials of knots (not links), we chose the same representation $T$ for each strand.  Matrices $\mathcal{R}_1$ and $\mathcal{R}_2$ represent crossings of the first and the second strands and the second and the third strands correspondingly. They are related to each other through a Racah mixing matrix (also known as $6j$-symbol and the Racah-Wigner coefficients) $\mathcal{U}$: $\mathcal{R}_2 = \mathcal{U} \mathcal{R}_1 \mathcal{U}^{\dagger}$.

By definition Racah matrices \cite{pent} relate the intertwiners:
\begin{equation}
\mathcal{ U}^{T_4}_{T_1, T_2, T_3}: ( (T_1 \otimes T_2) \otimes T_3 \rightarrow T_4 ) \longrightarrow ( T_1 \otimes  ( T_2 \otimes T_3) \rightarrow T_4 ).
\label{racahdef}
\end{equation}
i.e. describe deviations from the associativity in the product of representations.

There are different ways to calculate Racah matrices, such as via the pentagon relation \cite{pent, pent1, arbr2} or the highest wight vector method \cite{hwv, multeiv}. A number of  Racah matrices  were calculated with the highest weight vector method for different representations $T$ of $SU_q(N)$ \cite{knotebook}: for $T = [2,1]$ in \cite{r21}, for $T = [3,1]$ in \cite{r31} and for $T = [3,3]$ in \cite{r33}. These examples \cite{knotebook} enabled us to study certain properties of Racah matrices and therefore $\mathcal{R}_2$-matrices.

$\mathcal{R}$-matrices by definition satisfy the Yang-Baxter (YB) equation $\mathcal{R}_1 \mathcal{R}_2 \mathcal{R}_1 = \mathcal{R}_2 \mathcal{R}_1 \mathcal{R}_2$ (which is a mathematical form of the third Reidemeister move). It also relates $\mathcal{R}_1$ and $\mathcal{U}$ matrices:
\begin{equation}
\mathcal{R}_1 \mathcal{ U} \mathcal{R}_1 \mathcal{U}^{\dagger} \mathcal{R}_1 =\mathcal{ U} \mathcal{R}_1 \mathcal{U}^{\dagger} \mathcal{R}_1 \mathcal{ U} \mathcal{R}_1 \mathcal{U}^{\dagger}.
\label{YB2}
\end{equation}

Eigenvalues $\lambda_i$ of the $\mathcal{R}$-matrix \cite{eiv} are defined by irreducible representations $Q_i$ from $T^{\otimes 2} = \bigoplus_i a_i Q_i$.
\begin{equation}
\lambda_i = \epsilon_{Q_i}q^{\varkappa_{Q_i}},
\end{equation}
where $\varkappa_{Q_i} = \sum_{(m,n) \in Q_i} (m-n)$ is the image of the quadratic Casimir operator in the  representation $Q_i$, $\epsilon_{Q_i} = \pm 1$ depending whether $Q_i$ belongs to the symmetric or antisymmetric square of $T^{\otimes 2}$.

There are five possible sets of eigenvalues in $\mathcal{R}$-matrices:
\begin{enumerate}
	\item  all eigenvalues are different,
	\item some eigenvalues coincide because of the  multiplicity (when $a_i > 1 $ or/and $b_i > 1$ in recompositions $T^{\otimes 2} = \bigoplus_i a_i Q_i $ and $T^{\otimes 3} = \bigoplus_i b_i Q_i$) $\longrightarrow$ repetitive eigenvalues
	\item  some eigenvalues accidentally coincide: $\epsilon_{Q_i} q^{\varkappa_{Q_i}} = \epsilon_{Q_j} q^{\varkappa_{Q_j}}$, but $Q_i \neq Q_j$ $\longrightarrow$ accidentally coinciding eigenvalues,
	\item there are both repetitive ($\lambda$) and  accidentally coinciding eigenvalues ($\mu$), which don't coincide between each other ($ \nexists \ \lambda_i , \mu_j: \ \ \lambda_i = \mu_j$),
	\item there are both repetitive ($\lambda$) and  accidentally coinciding eigenvalues ($\mu$), which  also coincide ($\exists \ \lambda_i , \mu_j: \ \  \lambda_i = \mu_j $).
	
\end{enumerate}

In the case of the first set of eigenvalues, the equation (\ref{YB2}) can be solved for matrices up to $6 \times 6$ size, as it was done in \cite{eivcon}. As a result the eigenvalue hypothesis and its generalisation \cite{multeiv} were formulated. It claims that eigenvalues of $\mathcal{R}$-matrix fully define elements of the corresponding Racah matrix.

 It's also very important to consider other sets. In this paper we considered sets 2, 3 and 4.

The equation (\ref{trace}) has the following symmetry: one can arbitrary change the basis of $\mathcal{R}_1$- and $\mathcal{R}_2$-matrices: $\mathcal{R'}_1 = \mathcal{Q} \mathcal{R}_1 \mathcal{Q}^{\dagger}, \ \ \mathcal{R'}_2 = \mathcal{Q} \mathcal{R}_2 \mathcal{Q}^{\dagger}$ (one should apply the same transformation to the corresponding Racah matrix so that it can be used to calculate  HOMFLY polynomials). When $\mathcal{R} $-matrix contains coinciding eigenvalues, we can change just vectors corresponding to  them. Such rotation doesn't change $\mathcal{R}_1 $-matrix: 	$\mathcal{R'}_1 = \mathcal{O} \mathcal{R}_1 \mathcal{O}^{\dagger} = \mathcal{R}_1, \ \  \mathcal{R'}_2 = \mathcal{O} \mathcal{R}_2 \mathcal{O}^{\dagger}$. We used such rotations to obtain the results of this paper.

In experimental part of the present paper we rotated Racah matrices from \cite{knotebook} in the sectors of coinciding eigenvalues (repetitive and accidental) of $\mathcal{R} $-matrices. In case of repetitive eigenvalues such rotations represent just the change of the vectors in the corresponding representations. The resulting matrix still satisfies the definition (\ref{racahdef}). Rotations in the sectors corresponding to accidentally coinciding eigenvalues mix vectors from different irreducible representations. Thus the resulting matrix is no longer a Racah matrix. But it can still be used to calculate HOMFLY polynomials, because equation (\ref{trace}) is invariant under such rotations. This paper is devoted to studying the properties of Racah matrices and $\mathcal{R}_2$-matrices which come from these transformations.

The main results of the paper are two hypotheses.

\bigskip
\fbox{\begin{minipage}{15cm}
		\medskip
		\begin{enumerate}
			\item	$\mathcal{R}_2$-matrix can be transformed into a block diagonal form, if its eigenvalues satisfy two conditions: a number of pairs of accidentally coinciding eigenvalues $\geq N_a$, a number of pairs of multiple eigenvalues $\geq N_r$, where
			\begin{itemize}
				\item $N_a = 1$, $N_r = 0 $ or $N_a =0$, $N_r = 1 $ for matrices up to size $5 \times 5$,
				\item $N_a = 1$, $N_r = 1$ for $6 \times 6$ matrices,
				\item $N_a = 1$, $N_r = 2$ for $8 \times 8$.
			\end{itemize}

			\item The angle of the rotation in the sectors corresponding to accidentally coinciding eigenvalues from  the basis defined by the Racah matrix to the basis in which $\mathcal{R}_2$ is block-diagonal is $\pm \frac{\pi}{4}$.
		\end{enumerate}
		\smallskip
	\end{minipage}
	\rule{3mm}{0cm}}
\bigskip
\bigskip

The first hypothesis was proven for $\mathcal{R}_2$-matrices up to size $3 \times 3$. It can also be formulated for a Racah matrix which has a block-diagonal structure under certain transformations. In fact all calculations in experimental part were made for Racah matrices. The results can be found in the table \ref{tab}.

The resulting blocks can be found from the eigenvalues hypothesis since no eigenvalues in each block coincide. This is a very important for calculations fact and it follows from the first hypothesis.

The second hypothesis is the purely experimental observation. It is still an open question why the angle of the rotation in the ``accidental" sector is precisely $\pm \frac{\pi}{4} $.

\bigskip
The paper is organized as follows. In the section \ref{basics} there is the detailed information about $\mathcal{R}$-matrices and how they act on irreducible representations. In the section \ref{proof} we considered general case Racah matrices with one pair of coinciding eigenvalues. We straightforwardly solved the YB equation and showed that such $\mathcal{R}_2$-matrices have a block-diagonal structure. In the section \ref{calc} we described all rotations that were used in the experimental part. In sections \ref{s21} - \ref{s42} we listed all particular examples that were used to formulate the results of our work.

	\begin{table}[h!]
	\begin{center}
		\begin{tabular}{|c|c|c|c|c|c|c|c|}
			
			\hline
			
			size & repr &   \multicolumn{2}{|c|}{eigenvalues} & res  & blocks & block 1 & block 2\\
			\cline{3-4}
			& & repetitive & accidental & & & &\\

			\hline
			
			\multicolumn{8}{|c|}{\rule{0cm}{0.2cm}} \\
			\multicolumn{8}{|c|}{\textbf{[2,1]}} \\ [0.2cm]
			\hline
			
			\rule{0cm}{0.5cm}
			6 & [5,3,1] & $\lambda_1 = \lambda_2$ & $\lambda_3 = \lambda_4$ & + & (3,3) & $\underline{\lambda_1}$ ,$\lambda^*_3$, $\boxed{\lambda_6}$ &\underline{$\lambda_2$}, $\lambda^*_4$, $\boxed{\lambda_5}$  \\ [0.2cm]
	
			\hline
	
	\rule{0cm}{0.5cm}
			8 & [4,3,2] &$\lambda_4 = \lambda_5$, $\lambda_6 = \lambda_7$  & $\lambda_2 = \lambda_3$ & + & (3,5)  & $\lambda^*_3$,\underline{ $\lambda_5$},\underline{\underline{ $\lambda_6$}} &$\lambda_1$, $\lambda^*_2$,\underline{ $\lambda_4$}\\
			& & & & & &&  \underline{\underline{$\lambda_7$}}, $\lambda_8$  \\ [0.2cm]

			\hline
	\rule{0cm}{0.5cm}
			9 & [4,3,1,1] & $\lambda_4 = \lambda_5$, $\lambda_6 = \lambda_7$ &  $\lambda_2 = \lambda_3$ & -- & &&\\ [0.2cm]
	
			\hline

			\multicolumn{8}{|c|}{\rule{0cm}{0.2cm}} \\
		\multicolumn{8}{|c|}{\textbf{[3,1]}} \\ [0.2cm]
			
			\hline
		
	\rule{0cm}{0.5cm}		
			4 &  [4,3,2,2,1] & &  $\lambda_2 = \lambda_3 $ & +  & (1,3) & $\lambda^*_3$ &$\lambda_1$, $\lambda_2$, $\lambda^*_4$\\ [0.2cm]
			
			\hline
	\rule{0cm}{0.5cm}
			5 & [4,4,2,2] & & $\lambda_1 = \lambda_2 $ & + & (2,3) & $\lambda^*_1$, $\boxed{\lambda_4}$& $\lambda^*_2$, $\lambda_3$, $\boxed{\lambda_5}$ \\ [0.2cm]
			
			\hline
	\rule{0cm}{0.5cm}		
			6 & [4,3,3,1,1] & $\lambda_1 = \lambda_2 $ & $\lambda_3 = \lambda_4 $ &   +& (3,3)  & $\underline{\lambda_1}$, $\lambda_4^*$, $\boxed{\lambda_6}$ & $\underline{\lambda_2}$, $\lambda_3^*$, $\boxed{\lambda_5}$ \\ [0.2cm]
			
			\hline

	\rule{0cm}{0.5cm}		
			9 & [5,3,2,2] &  $\lambda_4 = \lambda_5 $ &  $\lambda_2 = \lambda_3 $ & -- & &&\\ [0.2cm]
			
			& [4,4,3,1] &  $\lambda_5 = \lambda_6 $,  $\lambda_7 = \lambda_8 $ &  $\lambda_2 = \lambda_3 $ & + & (3,6)  & $\lambda^*_3$, $\underline{\lambda_5}$, $\underline{\underline{\lambda_8}}$ &  $\lambda_1$, $\lambda^*_2$, $\lambda_4$\\ [0.2cm]
			& & & & & & & $\underline{\lambda_6}$, $\underline{\underline{\lambda_7}}$, $\lambda_9$  \\ [0.2cm]
			
			\hline

				\multicolumn{8}{|c|}{\rule{0cm}{0.2cm}} \\
			\multicolumn{8}{|c|}{\textbf{[3,2]}} \\ [0.2cm]

			\hline
			
			\rule{0cm}{0.5cm}
			6 & [4,3,3,2,2,1]	&$ \lambda_5 = \lambda_6 $ & $ \lambda_3 = \lambda_4 $ & + & (3,3)& $\boxed{\lambda_1}$, $\lambda^*_4$, $\underline{\lambda_5}$ & $\boxed{\lambda_2}$, $\lambda_3^*$, $\underline{\lambda_6}$\\ [0.2cm]
			& [6,4,2,1,1,1] &$ \lambda_1 = \lambda_2 $ & $ \lambda_3 = \lambda_4 $ & + & (3,3) & $\underline{\lambda_1}$, $\lambda^*_4$, $\boxed{\lambda_6}$& $\underline{\lambda_2}$, $\lambda^*_3$, $\boxed{\lambda_5}$ \\ [0.2cm]
			
			\hline

		\multicolumn{8}{|c|}{\rule{0cm}{0.2cm}} \\
	\multicolumn{8}{|c|}{\textbf{[4,1]}} \\ [0.2cm]
			
			\hline
\rule{0cm}{0.5cm}
			4 & [5,4,3,2,1]	& - & $ \lambda_2 = \lambda_3 $ & + & (1,3) & $\lambda^*_3$ 	&$\lambda_1$, $\lambda^*_2$, $\lambda_4$ \\ [0.2cm]

			\hline
\rule{0cm}{0.5cm}			
			6 & [5,5,3,2] &  $\lambda_3 = \lambda_4 $ &$ \lambda_5 = \lambda_6 $ & + & (3,3)&$\boxed{\lambda_1}$, $\underline{\lambda_3}$, $\lambda^*_5$  &$\boxed{\lambda_2}$, $\underline{\lambda_4}$, $\lambda^*_6$ \\ [0.2cm]
			&[5,4,4,1,1] & $\lambda_2 = \lambda_3 $, $ \lambda_4 = \lambda_5 $  & $ \lambda_6 = \lambda_7 $ & + & (3,3)&$\boxed{\lambda_1}$, $\lambda^*_3$, $\underline{\lambda_6}$ & $\boxed{\lambda_2}$, $\lambda^*_4$, $\underline{\lambda_5}$ \\ [0.2cm]
			
			& [6,4,2,2,1]	& - & $ \lambda_4 = \lambda_5 $ & -& &&\\ [0.2cm]
			
			\hline

			\multicolumn{8}{|c|}{\rule{0cm}{0.2cm}} \\
		\multicolumn{8}{|c|}{\textbf{[4,2]}} \\ [0.2cm]
			
			\hline
	\rule{0cm}{0.5cm}	
			4 & [5,4,3,3,2,1]	& - & $ \lambda_2 = \lambda_3 $ & + & (1,3)	 &$\lambda^*_3$ &$\lambda_1$, $\lambda^*_2$, $\lambda_4$ \\ [0.2cm]

			\hline
		
		\rule{0cm}{0.5cm}	
			5 & [5,5,3,3,1,1] & - & $\lambda_4 = \lambda_5 $ & + & (2,3)  & $\boxed{\lambda_1}$, $\lambda^*_4$ &$\boxed{\lambda_2}$, $\lambda_3$, $\lambda^*_5$  \\ [0.2cm]
			\hline
			
	\rule{0cm}{0.5cm}	
			6 & [5,4,4,2,2,1]	&  $\lambda_5 = \lambda_6 $ & $ \lambda_3 = \lambda_4 $ & + & (3,3) 	& $\boxed{\lambda_1}$, $\lambda^*_3$, $\underline{\lambda_6}$ &$\boxed{\lambda_2}$, $\lambda^*_4$, $\underline{\lambda_5}$ \\ [0.2cm]
			& [11,6,1]	&  $\lambda_1 = \lambda_2 $ & $ \lambda_3 = \lambda_4 $ & + & (3,3) 	&$\underline{\lambda_1}$, $\lambda^*_3$, $\boxed{\lambda_5}$  & $\underline{\lambda_2}$, $\lambda^*_4$, $\boxed{\lambda_6}$\\ [0.2cm]
			\hline

		\end{tabular}
	\end{center}
	\label{tab}
	\caption{\textbf{Matrices with accidentally coinciding eigenvalues.}
	In the last two columns accidentally coinciding eigenvalues are marked with stars, repetitive eigenvalues are underlined (and twice underlined),``symmetric -- antisymmetric'' eigenvalues  are boxed.}
\end{table}

\section{$\mathcal{R}$-matrices in three-strand braids \label{basics}}

$\mathcal{R}$-matrix is an operator which acts on a tensor product of two representations $T_i \otimes T_j$ ($T_k$ is a representation of $SU_q(N)$) which are associated with lines of a braid. As it was mentioned earlier in the case of three strand braids and identical representations there are two $\mathcal{R}$-matrices which are used in the RT formalism:

\begin{eqnarray}
\mathcal{R}_1 = \mathcal{R} \otimes I \in Mat(T \otimes T \otimes T), \\
\mathcal{R}_2 = I \otimes \mathcal{R} \in Mat(T \otimes T \otimes T),
\end{eqnarray}
where $I$ is the unit operator.

In general case vectors of irreducible representations $Q_k$  ($T_i \otimes T_j = \bigoplus_k a_k Q_k $) are eigenvectors of $\mathcal{R}$-matrix. It follows from the fact that $\mathcal{R}$-matrix commutes with the coproduct.
To show this fact, let's consider how $\mathcal{R}$-matrix acts on vectors $v_i$ and $u_j$ of two irreducible representations $Q_1$ and $Q_2$ (${\rm dim}(Q_1) < {\rm dim}(Q_2)$). $v_1$ and $u_1$ are the highest weight vectors of representations $Q_1$ and $Q_2$, $T^+$ and $T^-$ are the raising and the lowering operators.

\begin{equation}
\begin{array}{l}
\mathcal{R} v_1 = \sum_{i = 1}^{{\rm dim} (Q_1)} a_i v_i + \sum_{j =1}^{{\rm dim} (Q_2)} b_j u_j, \\
T^+ \mathcal{R} v_1 =  \sum_{i = 2}^{{\rm dim} (Q_1)} a_i v_{i-1} + \sum_{j =2}^{{\rm dim} (Q_2)} b_j u_{j-1}
\end{array}
\end{equation}
From the other point of view
\begin{equation}
\begin{array}{l}
T^+ \mathcal{R} v_1 =  \mathcal{R} T^+ v_1 = 0,
\end{array}
\end{equation}
it means that  $a_i = b_j = 0$ for $i = 2, \dots , {\rm dim}(Q_1), j = 1, \dots,{\rm dim}(Q_2) $ and now $\mathcal{R}v_1 = a_1 v_1+b_1 u_1$.
\begin{equation}
\begin{array}{l}
(T^-)^{{\rm dim}(Q_1)} \mathcal{R} v_1 = b_1 u_{{\rm dim}(Q_1)+1} = \mathcal{R} (T^-)^{{\rm dim}(Q_1)}  v_1 = 0 \Rightarrow
b_1 = 0 \Rightarrow
\mathcal{R} v_1 = a_1 v_1
\end{array}
\end{equation}
Acting on this expression with the raising operator one can show that
\begin{equation}
\begin{array}{l}
\mathcal{R} v_i = a_1 v_i.
\end{array}
\end{equation}
One can use the same procedure to show that it's true for each irreducible representation.

\begin{equation}
\mathcal{R} Q_i = \lambda_i Q_i = \pm q^{\varkappa_i} Q_i
\end{equation}

That's why there is a basis where $\mathcal{R}_1$ is diagonal.
\begin{equation}
\mathcal{R}_1 (T_1 \otimes T_2 \otimes T_3) = \mathcal{R} (T_1 \otimes T_2) \otimes I(T_3) = \mathcal{R} \bigoplus_i Q_i \otimes T_3 =  \bigoplus_i \lambda_i Q_i \otimes T_3.
\end{equation}

\begin{equation}
T_1 \otimes T_2 \otimes T_3 = \bigoplus_k a_k Q'_k.
\end{equation}

$\mathcal{R}_1$ and $\mathcal{R}_2$ have block-diagonal forms with blocks $\mathcal{R}_{1;Q'_k}$ and $\mathcal{R}_{2;Q'_k}$:

\begin{equation}
\begin{array}{l}
\mathcal{R}_{1;Q'_k} = {\rm diag}(\lambda_{j_1}, \dots \lambda_{j_{a_j}}),  \\
\mathcal{R}_{2;Q'_k} = (\mathcal{U}_{Q'_k})^{\dagger} \mathcal{R}_{1;Q'_k} \mathcal{U}_{Q'_k},
\end{array}
\end{equation}
where $\lambda_{j_l}$ are  eigenvalues  of representations $Q_{j_l}$ (from the  sum $T_1 \otimes T_2 = \bigoplus_i Q_i$: $Q_{j_l} \otimes T_3 = \bigoplus_m Q'_m$  and in this sum there is $Q'_{m_l} = Q'_k$).
The equation (\ref{trace}) is also modified \cite{hwv}:
\begin{equation}
H_T^{(m_1, n_1 | m_2, n_2 | \dots)} = \sum \limits_{Q \in T^{\otimes 3}} S^*_{Q} \Tr _Q (\mathcal{R}_{1;Q}^{m_1} \mathcal{ U}_Q \mathcal{R}_{1;Q}^{n_1} \mathcal{ U}_Q^{\dagger} \mathcal{R}_{1;Q}^{m_2} \mathcal{ U}_Q \mathcal{R}_{1;Q}^{n_2} \mathcal{ U}_Q^{\dagger} \dots),
\end{equation}
where $m_i$ and $n_i$ enumerate the crossings in the braid,  $S^*_{Q}$ are the Schur polynomials\footnote{there also exists a constructive hook formula, which allows one to calculate them easily:
	\begin{equation}
	S^*_Q(A,q)=\prod_{(i,j)\in Q}
	\frac{Aq^{i-j}-A^{-1}q^{j-i}}{q^{h_{i,j}}-q^{-h_{i,j}}},
	\begin{picture}(105,15)(-35,-15)
	\put(0,0){\line(1,0){70}}
	\put(0,-10){\line(1,0){70}}
	\put(0,-20){\line(1,0){60}}
	\put(0,-30){\line(1,0){40}}
	\put(0,-40){\line(1,0){20}}
	\put(0,-50){\line(1,0){20}}
	\put(0,0){\line(0,-1){50}}
	\put(10,0){\line(0,-1){50}}
	\put(20,0){\line(0,-1){50}}
	\put(30,0){\line(0,-1){30}}
	\put(40,0){\line(0,-1){30}}
	\put(50,0){\line(0,-1){20}}
	\put(60,0){\line(0,-1){20}}
	\put(70,0){\line(0,-1){10}}
	\put(15,-15){\makebox(0,0)[cc]{\textbf{x}}}
	\put(15,5){\makebox(0,0)[cc]{$i$}}
	\put(-5,-15){\makebox(0,0)[cc]{$j$}}
	\qbezier(19,-11)(45,20)(55,-15)
	\put(40,10){\makebox(0,0)[cc]{$k$}}
	\qbezier(11,-19)(-17,-40)(15,-45)
	\put(60,-40){\makebox(0,0)[lc]{$h_{i,j}=k+l+1$.}}
	\end{picture}
	\end{equation}
	\noindent $[n]_q$ denotes the quantum number $n$, i.e., $[n]_q\equiv\frac{q^n-q^{-n}}{q-q^{-1}}$.
}.

\section{The proof of the first hypothesis \label{proof}}

In the experimental part  we have considered Racah matrices up to six dimensions. Among all the examples we encountered just cases with one pair of accidentally coinciding eigenvalues. That's why, we'll consider now a general case of $\mathcal{R}$-matrices with a pair of coinciding eigenvalues.
\begin{enumerate}
	\item \textbf{$n = 2$}
	
	In this case
	
	\begin{equation}
	\mathcal{R} =
	\begin{pmatrix}
	\lambda & \\
	& \lambda \\
	\end{pmatrix}
	=
	\lambda
	\begin{pmatrix}
	1 & \\
	& 1 \\
	\end{pmatrix},
	\end{equation}
it means that $\mathcal{R}_2 $ is proportional to the identity matrix too, because $\mathcal{R}_2 = U \mathcal{R} U^{\dagger}$.
	
	\item \textbf{ $n = 3$}
	
	Here
	
	\begin{equation}
	\mathcal{R} =
	\begin{pmatrix}
	\lambda_1 & & \\
	& \lambda_1 & \\
	& & \lambda_3 \\
	\end{pmatrix}
	\end{equation}
	$\mathcal{R}_2$ is symmetric because $\mathcal{R}$ is diagonal and $\mathcal{ U}$ is real.
	
	\begin{equation}
	\mathcal{R}_2 =
	\begin{pmatrix}
	\alpha_1 & \beta_1 & \gamma_1 \\
	\beta_1 & \beta_2 & \gamma_2 \\
	\gamma_1 & \gamma_2 & \gamma_3 \\
	\end{pmatrix}
	\end{equation}	
	
	Since one can always use a rotation to make $\beta_1 = 0$:
	
		\begin{equation}
	\mathcal{O}_{(\beta_1 = 0)} = \begin{pmatrix}
	c & -s & \\
	s & c & \\
	& & 1 \\
	\end{pmatrix},
	\end{equation}
where
\begin{equation}
\begin{array}{ll}
	c =  \frac{1}{\sqrt{\frac{\left(\sqrt{(\alpha _1-\beta _2)^2+4 \beta _1^2}+\alpha _1-\beta _2\right)^2}{4 \beta _1^2}+1}}, &
	s = \frac{\sqrt{(\alpha _1-\beta _2)^2+4 \beta _1^2}+\alpha _1-\beta _2}{2 \beta _1 \sqrt{\frac{\left(\sqrt{(\alpha _1-\beta _2)^2+4
					\beta _1^2}+\alpha _1-\beta _2\right)^2}{4 \beta _1^2}+1}}, \\
\end{array}
\end{equation}
it's reasonable to consider $\tilde{\mathcal{R}}_2 = 	\mathcal{O}_{(\beta_1 = 0)} \mathcal{R}_2 	\mathcal{O}_{(\beta_1 = 0)}^{\dagger} $. Such rotation doesn't change $\mathcal{R}_1$-matrix because it's in the sector of coinciding eigenvalues.

Renaming elements we get
\begin{equation}
\tilde{\mathcal{R}}_2 =
\begin{pmatrix}
a_1 & 0 & c_1 \\
0 & a_2 & c_2 \\
c_1 & c_2 & c_3 \\
\end{pmatrix}
\end{equation}

$\mathcal{R}$ and $\tilde{\mathcal{R}}_2$ obey the Yang-Baxter equation (YB). The element ${\rm YB}_{12}$ of this equation is
\begin{equation}
\lambda_3 c_1 c_2 = 0.
\end{equation}
There are two possible solutions ($c_1 = 0$ and $c_2= 0$) and both of them make 	$\tilde{\mathcal{R}}_2$ block-diagonal.
\end{enumerate}

\section{Calculations \label{calc}}
We studied a number of $\mathcal{R}$-matrices in different representations of $SU_q(N)$ and picked up matrices with coinciding eigenvalues. Then we rotated corresponding Racah matrices \cite{knotebook} and found out that almost all matrices rotated in a sector of accidentally coinciding eigenvalues became block diagonal, therefore corresponding  $\mathcal{R}_2$-matrices should be block-diagonal too. One can see the results in the table \ref{tab}.
If $\mathcal{R} $-matrix has coinciding eigenvalues ($\lambda_i = \lambda_j$)
\begin{equation}
\mathcal{R} =
\begin{pmatrix}
\ddots & & & & \\
 & \lambda_i & & & \\
 & & \ddots & & \\
 & & & \lambda_j &  \\
 & & & & \ddots
\end{pmatrix}
\end{equation}
 we can rotate corresponding Racah matrix $\mathcal{ U}$ in the sector $i-j$ with the rotation matrix $\mathcal{O}$:
\begin{equation}
\mathcal{O} =
\begin{pmatrix}
\ddots & & & & \\
& c & &  -s  &\\
& & \ddots & &\\
& s & &  c &  \\
 & & & & \ddots
\end{pmatrix}
\label{rot}
\end{equation}

\begin{equation}
\mathcal{U}' = \mathcal{O} \mathcal{U} \mathcal{O}^T,
\end{equation}

\begin{equation}
\begin{pmatrix}
c & 0 & \cdots & 0 & -s \\
0 & 1 & \cdots & 0 & 0 \\
& &  \ddots & & \\
0 & 0 & \cdots & 1 & 0 \\
s & 0 & \cdots & 0 & c \\
\end{pmatrix}
\begin{pmatrix}
u_{i,i} &u_{i, i+1}& \cdots & u_{i, j-1} & u_{i,j} \\
u_{i+1, i} & u_{i+1,i+1} & \cdots & u_{i+1, j-1} & u_{i+1, j} \\
 & & \ddots & & \\
u_{j-1, i} & u_{j-1, i+1} & \cdots & u_{j-1, j-1} & u_{j-1, j} \\
u_{j, i} & u_{j,i+1} & \cdots & u_{j, j-1} &  u_{j, j} \\
\end{pmatrix}
\begin{pmatrix}
c & 0 & \dots & 0 & s \\
0 & 1 & \dots & 0 & 0 \\
& &  \ddots & & \\
0 & 0 & \dots & 1 & 0 \\
-s & 0 & \dots & 0 & c \\
\end{pmatrix} =
\end{equation}

\begin{equation}
\begin{pmatrix}
u'_{i,i} &u'_{i, i+1}& \cdots & u'_{i, j-1} & u'_{i,j} \\
u'_{i+1, i} & u_{i+1,i+1} & \cdots & u_{i+1, j-1} & u'_{i+1, j} \\
& & \ddots & & \\
u'_{j-1, i} & u_{j-1, i+1} & \cdots & u_{j-1, j-1} & u'_{j-1, j} \\
u'_{j, i} & u'_{j,i+1} & \cdots & u'_{j, j-1} &  u'_{j, j} \\
\end{pmatrix}.
\end{equation}
To make $\mathcal{U}'$ block-diagonal one can determine the angle of rotation from the requirement that diagonal elements inside the sector of final matrix equal zero:
\begin{equation}
u'_{i, j} = u'_{j, i} = 0.
\label{req}
\end{equation}
This method works only if $u_{i,j} = u_{j,i}$ in the initial matrix, otherwise equations are unsolvable. If  $u_{i,j} \neq u_{j,i}$   we can use two different rotation matrices:

\begin{equation}
\begin{array}{ccc}
\mathcal{O}_1 =
\begin{pmatrix}
\ddots & & & & \\
& c_1 & &  -s_1  &\\
& & \ddots & &\\
& s_1 & &  c_1 &  \\
& & & & \ddots
\end{pmatrix} &
{\rm and} &
\mathcal{O}_2 =
\begin{pmatrix}
\ddots & & & & \\
& c_2 & &  -s_2  &\\
& & \ddots & &\\
& s_2 & &  c_2 &  \\
& & & & \ddots
\end{pmatrix} \\
\end{array}
\end{equation}
 to get a new mixing matrix:
\begin{equation}
\mathcal{U}' = \mathcal{O}_1 \mathcal{U} \mathcal{O}_2^T,
\end{equation}

\begin{equation}
\label{rrot}
\begin{pmatrix}
c_1 & 0 & \cdots & 0 & -s_1 \\
0 & 1 & \cdots & 0 & 0 \\
& &  \ddots & & \\
0 & 0 & \cdots & 1 & 0 \\
s_1 & 0 & \cdots & 0 & c_1 \\
\end{pmatrix}
\begin{pmatrix}
u_{i,i} &u_{i, i+1}& \cdots & u_{i, j-1} & u_{i,j} \\
u_{i+1, i} & u_{i+1,i+1} & \cdots & u_{i+1, j-1} & u_{i+1, j} \\
& & \ddots & & \\
u_{j-1, i} & u_{j-1, i+1} & \cdots & u_{j-1, j-1} & u_{j-1, j} \\
u_{j, i} & u_{j,i+1} & \cdots & u_{j, j-1} &  u_{j, j} \\
\end{pmatrix}
\begin{pmatrix}
c_2 & 0 & \dots & 0 & s_2 \\
0 & 1 & \dots & 0 & 0 \\
& &  \ddots & & \\
0 & 0 & \dots & 1 & 0 \\
-s_2 & 0 & \dots & 0 & c_2 \\
\end{pmatrix} =
\end{equation}
\begin{equation}
\begin{pmatrix}
u'_{i,i} &u_{i, i+1}& \cdots & u'_{i, j-1} & u'_{i,j} \\
u'_{i+1, i} & u_{i+1,i+1} & \cdots & u_{i+1, j-1} & u'_{i+1, j} \\
& & \ddots & & \\
u'_{j-1, i} & u_{j-1, i+1} & \cdots & u_{j-1, j-1} & u'_{j-1, j} \\
u'_{j, i} & u'_{j,i+1} & \cdots & u'_{j, j-1} &  u'_{j, j} \\
\end{pmatrix}.
\end{equation}
The angle of the rotation can be found from the same requirement (\ref{req}) as in the previous case.

\section{Representation [2,1] \label{s21}}
We know that
\begin{equation}
[2,1] \otimes [2,1] = [4,2] \oplus \underline{[4,1,1]} \oplus \underline{[3,3]} \oplus [3,2,1] \oplus \underline{[3,2,1]} \oplus [3,1,1,1] \oplus [2,2,2] \oplus \underline{[2,2,1,1]},
\end{equation}
where there are two types of irreducible representations: symmetric and antisymmetric (underlined).
Corresponding $\mathcal{R}$-matrices have eigenvalues:
\begin{equation}
\begin{array}{cccc}
\lambda_{[4,2]} = \frac{1}{q^5}, &
\lambda_{[4,1,1]} = -\frac{1}{q^3}, &
\lambda_{[3,3]} =  -\frac{1}{q^3}, &
\lambda_{[3,2,1]_{\pm}}  = \pm 1, \\
\lambda_{[3,1,1,1]} = q^3, &
\lambda_{[2,2,2]} = q^3, &
\lambda_{[2,2,1,1]} = -q^5, & \\
\end{array}
\end{equation}
and one can see that there are two pairs of accidentally coinciding eigenvalues:
\begin{equation}
\begin{array}{c}
\lambda_{[4,1,1]} = \lambda_{[3,3]} =  -\frac{1}{q^3}, \\
\lambda_{[3,1,1,1]} = \lambda_{[2,2,2]} = q^3.

\end{array}
\end{equation}
\begin{table}[h]
	\begin{center}
		\begin{tabular}{|c|c|c|c|c|c|}
			\hline
			\multicolumn{6}{|c|}{\rule{0cm}{0.2cm}} \\
			\multicolumn{6}{|c|}{Table of matrices with accidentally coinciding eigenvalues} \\
			\multicolumn{6}{|c|}{$ [2,1]$} \\ [0.2cm]
			
			\hline
			
	     	size & repr &   \multicolumn{2}{|c|}{eigenvalues} & result  & blocks \\
	     	\cline{3-4}
			& & repetitive & accidental & & \\
		
 \hline
	\rule{0cm}{0.5cm}
 6 & [5,3,1] & $\lambda_1 = \lambda_2$ & $\lambda_3 = \lambda_4$ & + & (3,3)\\ [0.2cm]
 \hline

	\rule{0cm}{0.5cm}
8 & [4,3,2] &$\lambda_4 = \lambda_5$, $\lambda_6 = \lambda_7$  & $\lambda_2 = \lambda_3$ & + &(3,5) \\ [0.2cm]
\hline

	\rule{0cm}{0.5cm}
9 & [4,3,1,1] & $\lambda_4 = \lambda_5$, $\lambda_6 = \lambda_7$ &  $\lambda_2 = \lambda_3$ & - & \\ [0.2cm]
\hline

\end{tabular}
\end{center}
\end{table}
\subsection{[5,3,1]}

\begin{equation}
\begin{array}{c}
\mathcal{R}_{[5,3,1]} = {\rm diag} (q^{\varkappa_{[2,2,1,1]}}, \  q^{\varkappa_{[2,2,1,1]}}, \ -q^{\varkappa_{[3,1,1,1]}}, \  -q^{\varkappa_{[2,2,2]}}, \  -q^{\varkappa_{[3,2,1]}}, \  q^{\varkappa_{[3,2,1]}}) = \\
{\rm diag}  (-q^{-5}, \ -q^{-5}, \ q^{-3}, \  q^{-3}, \ -1, \ 1). \\
\end{array}
\end{equation}

We see that $\lambda_1 = \lambda_2$ and $\lambda_3 = \lambda_4$, that's why we rotate $\mathcal{U}_{[5,3,1]}$ in sectors $1-2$ and $3-4$. And we have

\begin{equation}
\begin{array}{l}
c_{12}^{[5,3,1]} =\frac{1}{\sqrt{2}} \sqrt{1 + \frac{q (1 + q^2) (1 + q^4)^2}{(1 + q^2 (1 + (-1 + q) q) (1 + q + q^2) (2 + q^4 + q^6)}}, \\ \\
s_{12}^{[5,3,1]} = \frac{1}{\sqrt{2}} \sqrt{\frac{(1 + (-1 + q) q) (1 + q + q^2 + q^3 + q^4) (1 + (-1 + q) q (1 + (-1 + q) q) (1 + q + q^2))}{(1 +
q^2 (1 + (-1 + q) q) (1 + q + q^2) (2 + q^4 + q^6)}},
\end{array}
\end{equation}
and
\begin{equation}
\boxed{\begin{array}{ll}
c_{34}^{[5,3,1]} = \frac{1}{\sqrt{2}}, & s_{34}^{[5,3,1]} = -\frac{1}{\sqrt{2}}.
\end{array}}
\end{equation}
After the rotations $\mathcal{U}_{[5,3,1]}$ becomes block-diagonal:
\begin{equation}
 \mathcal{U'}_{[5,3,1]} =
 \begin{pmatrix}
 -\frac{q^2 (1 + q + q^2) }{(1 + q^2) y_1} & 0 &  x_{13} & 0 & 0 & x_ {16} \\

 0 &  \frac{q^2 (1 - q + q^2)}{(1 + q^2) y_ 2} & 0 & x_ {24} & x_ {25} & 0 \\
 x_ {13} &  0 &  \frac{-1 + q - q^2 + q^3 - q^4}{(1 + q^2) (1 - q + q^2)}&  0 & 0 &  x_ {36} \\

 0 & x_ {24} &  0 & \frac{ y1}{y1 + q^2} & x_ {45} & 0 \\

 0 & x_ {52} &  0 &  x_ {45} & -\frac{q^3}{1 + q^2 - q^3 + q^4 + q^6} & 0 \\
 x_ {16} &  0 &  x_ {36} &  0 & 0 &  -\frac{q^3}{y1 + q^6} \\
\end{pmatrix},
\label{m531}
\end{equation}
where
\begin{equation}
\begin{array}{ll}
x_ {13} = \frac{q}{1 + q^2} \sqrt{\frac{1 + q + q^2 + q^3 + q^4 + q^5 + q^6}{1 + q^2 + q^3 + q^4 + q^6}}, &

x_ {24} = -\frac{q}{1 + q^2} \sqrt{\frac{1 - q + q^2 - q^3 + q^4 - q^5 + q^6}{1 + q^2 - q^3 + q^4 + q^6} }, \\ \\
x_ {16} = \frac{(1 + q^4) \sqrt{\frac{1 + q + q^2 + q^3 + q^4 + q^5 + q^6}{1 - q + q^2 + q^4 - q^5 + q^6}}}{1 + q + q^2 + q^3 + q^4} , &

x_ {45} = \frac{q \sqrt{\frac{1 + q^4}{1 - q + q^2 - q^3 + q^4}}}{1 + q + q^2} , \\ \\

x_ {36} = \frac{q \sqrt{ \frac{1 + q^4}{1 + q + q^2 + q^3 + q^4} } }{1 - q + q^2} , &

x_ {25} = \frac{\sqrt{(1 + q^4) (1 + q^2 - q^3 + q^4 + q^6) (1 - q + q^2 - q^3 + q^4 - q^5 + q^6)}}{(1 + q + q^2) (1 - q + q^2 - q^3 + q^4)^{3/2} },  \\ \\

x_ {52} = \frac{(1 + q^4) \sqrt{ \frac{1 - q + q^2 - q^3 + q^4 - q^5 + q^6}{1 + q + q^2 + q^4 + q^5 + q^6} }}{1 - q + q^2 - q^3 + q^4},  & \\ \\

y_ 1 = 1 + q + q^2 + q^3 + q^4,  &

y_ 2 = 1 - q + q^2 - q^3 + q^4. \\
\end{array}
\end{equation}
Changing lines of the matrix (\ref{m531}) we can see that it has a block-diagonal structure with blocks $3 \times 3$.

\subsection{[4,3,2]}

\begin{equation}
\begin{array}{c}
\mathcal{R}_{[4,3,2]} = {\rm diag} \ (q^{ \varkappa_{[2,2,1,1]}}, \ -q^{\varkappa_{[3,1,1,1]}}, \ -q^{\varkappa_{[2,2,2]}}, \ -q^{\varkappa_{[3,2,1]}}, \ -q^{\varkappa_{[3,2,1]}}, \ q^{\varkappa_{[3,2,1]}}, \ q^{\varkappa_{[3,2,1]}}, \ q^{\varkappa_{[3,3]}}  ) = \\
{\rm diag} \ ( -q^{5}, \ q^{3}, \ q^{3}, \ -1, \ -1, \ 1, \ 1, \ -q^{-3})
\end{array}
\end{equation}

Despite the fact that here we have tree pairs of coinciding eigenvalues: $\lambda_2 = \lambda_3, \ \lambda_4 = \lambda_5$ and $\lambda_6 = \lambda_7$, we need \underline{just one rotation} in sector $2-3$ with

\begin{equation}
\boxed{\begin{array}{ll}
	c_{23}^{[4,3,2]} = \frac{1}{\sqrt{2}}, & s_{23}^{[4,3,2]} = -\frac{1}{\sqrt{2}}.
	\end{array}}
\end{equation}
to make $\mathcal{U}_{[5,3,1]}$ block-diagonal:
\begin{equation}
\mathcal{U'}_{[4,3,2]} =
\begin{pmatrix}
x_{11} & x_{12} & 0 &  x_{14} &  0 &  0 & x_{17} &  x_{18} \\
x_{12} & \frac{2 q^2}{(1 + q^2)^2} & 0 & x_{24} & 0 & 0 & x_{27}& x_{28} \\
0 & 0 & 0 & 0 & -\frac{1}{\sqrt{2}} &-\frac{1}{\sqrt{2}} & 0 & 0 \\
x_{14} & x_{24} & 0 & \frac{1 + q - 3 q^2 + q^3 + q^4}{
		2 (1 - q + q^2 - q^3 + q^4)} & 0 &  0 & \frac{1}{2} & -\frac{q^2}{(1 + q^2) \sqrt{1 + q^4}} \\
0 & 0 &  -\frac{1}{\sqrt{2}} & 0 & \frac{1}{2} & -\frac{1}{2} & 0 & 0 \\
0 & 0 & -\frac{1}{\sqrt{2}} & 0 & -\frac{1}{2} & \frac{1}{2} & 0 & 0 \\
x_{17} & x_{27} & 0 & \frac{1}{2} & 0 & 0 & \frac{\sqrt{(-1 + q + 3 q^2 + q^3 - q^4)^2}}{2 (1 + q + q^2 + q^3 + q^4)} &  -\frac{q^2}{(1 + q^2) \sqrt{1 + q^4}} \\
x_{18} & x_{28} & 0 &  -\frac{q^2}{ (1 + q^2) \sqrt{1 + q^4} } & 0 & 0 & -\frac{q^2}{ (1+ q^2) \sqrt{1 + q^4} } & \frac{q^4}{(1 + q^2)^2 (1 + q^4)} \\
\end{pmatrix},
\end{equation}
where
\begin{equation}
\begin{array}{ll}
x_{11} = \frac{q^4 (1 + q^2 + 3 q^4 + q^6 + q^8)}{(1 + q^2)^2 (1 + q^2 +
2 q^4 + 2 q^6 + 2 q^8 + q^{10} + q^{12})}, &
x_{12} =  \sqrt{\frac{
2q^6 (1 + q^4 + q^8)}{(1 + q^2)^4 (1 + q^2 + 2 q^4 + 2 q^6 + 2 q^8 +
q^{10} + q^{12})}}, \\ \\
x_{14} = \sqrt{\frac{
q^2 (1 - q + q^2)^3 (1 + q + q^2) (1 - q^2 + q^4)}{(1 + q^2)^2 (1 +
q^4) (1 - q + q^2 - q^3 + q^4)^2}}, &
x_{17} = -\sqrt{\frac{q^2 (1 - q + q^2) (1 + q + q^2)^3 (1 - q^2 + q^4)}{(1 + q^2)^2 (1 + q^4) (1 + q + q^2 + q^3 + q^4)^2}} ,\\ \\
x_{24} = \sqrt{\frac{(q-1)^4 (1 + q + q^2 + q^3 + q^4)}{2(1 + q^2)^2 (1-q + q^2 - q^3 + q^4)}}, &
x_{27} = -\sqrt{\frac{(1 + q)^4 (1 - q + q^2 - q^3 + q^4)}{2(1 + q^2)^2 (1 +
q + q^2 + q^3 + q^4)}} ,\\ \\
x_{18} = \sqrt{\frac{(1 - q^2 + q^4) (1 + q^2 + q^4)^3}{(1 + q^2)^4 (1 +
q^4)^2}} ,&
x_{28} = -\frac{\sqrt{2} q \sqrt{\frac{1 + q^2 + q^4 + q^6 + q^8}{
1 + q^4}}}{(1 + q^2)^2}. \\
\end{array}
\label{m432}
\end{equation}
Changing lines of the matrix (\ref{m432}) we can see that it  has a block-diagonal structure with blocks $5 \times 5$ and $3 \times 3$.

\section{Representation [3,1] \label{s31}}
\begin{equation}
[3,1]^{\otimes 2} = [6,2] \oplus[6,1,1]\oplus[5,3]\oplus2[5,2,1]\oplus[5,1,1,1]\oplus[4,4]\oplus2[4,3,1]\oplus[4,2,2]\oplus[4,2,1,1]\oplus[3,3,2]\oplus[3,3,1,1].
\end{equation}
Eigenvalues are:
\begin{equation}
\begin{array}{llll}
 \lambda_{[6,2]} = q^{-14}, &
 \lambda_{[6,1,1]} = -q^{-12}, &
 \lambda_{[5,3]} = -q^{-10}, &
 \lambda_{[5,2,1]_{\pm}} = \pm q^{-7}, \\
 \lambda_{[5,1,1,1]} = q^{-4}, &
 \lambda_{[4,4]} = q^{-8}, &
 \lambda_{[4,3,1]_{\pm}} = \pm q^{-4}, &
 \lambda_{[4,2,2]} = q^{-2}, \\
 \lambda_{[4,2,1,1]} = -1, &
 \lambda_{[3,3,2]} = -1, &
 \lambda_{[3,3,1,1]} = q^{2}.

\end{array}
\end{equation}

And here we have two pairs of eigenvalues which coincide:
\begin{equation}
\begin{array}{c}
\lambda_{[5,1,1,1]} = \lambda_{[4,3,1]_+} = q^{-4} \\
\lambda_{[4,2,1,1]} = \lambda_{[3,3,2]} = -1 \\
\end{array}
\end{equation}

\begin{table}[h!]
\begin{center}
\begin{tabular}{|c|c|c|c|c|c|}
	\hline
	
	\multicolumn{6}{|c|}{\rule{0cm}{0.2cm}} \\
\multicolumn{6}{|c|}{Table of matrices with accidentally coinciding eigenvalues} \\
\multicolumn{6}{|c|}{$ [3,1]$} \\ [0.2cm]

\hline

size & repr &   \multicolumn{2}{|c|}{eigenvalues} & result  & blocks \\
\cline{3-4}
& & repetitive & accidental & & \\

\hline
\rule{0cm}{0.5cm}
4 &  [4,3,2,2,1] & &  $\lambda_2 = \lambda_3 $ & + & $(3,1)$ \\ [0.2cm]

\hline
\rule{0cm}{0.5cm}
5 & [4,4,2,2] & & $\lambda_1 = \lambda_2 $ & + & $(3,2)$\\ [0.2cm]

\hline
\rule{0cm}{0.5cm}

6 & [4,3,3,1,1] & $\lambda_1 = \lambda_2 $ & $\lambda_3 = \lambda_4 $ &   + & $(3,3)$ \\ [0.2cm]

\hline
\rule{0cm}{0.5cm}
9 & [5,3,2,2] &  $\lambda_4 = \lambda_5 $ &  $\lambda_2 = \lambda_3 $ & - & -\\ [0.2cm]
& [4,4,3,1] &  $\lambda_5 = \lambda_6 $,  $\lambda_7 = \lambda_8 $ &  $\lambda_2 = \lambda_3 $ & + & $(3,6)$ \\ [0.2cm]

 \hline
\end{tabular}
\end{center}
\end{table}

\subsection{[4,3,2,2,1]}
\begin{equation}
\mathcal{R}_{[4,3,2,2,1]} = {\rm diag}(q^{\varkappa_{[3,3,1,1]}} , \ -q^{\varkappa_{[4,2,1,1]}}, \ -q^{\varkappa_{[3,3,2]}} , \-q^{\varkappa_{[4,2,2]}} )
=
{\rm diag}(q^{2}, \-1 , \-1 , \ q^{-2} ).
\end{equation}
After the rotation in the sector $2 - 3$ with
\begin{equation}
\boxed{\begin{array}{ll}
	c_{23}^{[4,3,2,2,1]} = \frac{1}{\sqrt{2}}, & s_{23}^{[4,3,2,2,1]} = -\frac{1}{\sqrt{2}}.
	\end{array}}
\end{equation}
 $\mathcal{R}_{[4,3,2,2,1]}$ becomes block-diagonal:

\begin{equation}
\mathcal{U'}_{[4,3,2,2,1]} =\mathcal{O}_{23}^{[4,3,2,2,1]} \mathcal{U}_{[4,3,2,2,1]} ( \mathcal{O}_{23}^{[4,3,2,2,1]} ) ^{\dagger} =
  \begin{pmatrix}
  \frac{q^2}{(1 + q^2)^2} & \frac{ (\sqrt{2} q \sqrt{1 + q^2 + q^4}}{(1 + q^2)^2} & 0 & \frac{1 + q^2 + q^4}{(1 + q^2)^2} \\
  	-\frac{\sqrt{2} q \sqrt{1 + q^2 + q^4}}{(1 + q^2)^2} & \frac{(-1 - q^4)}{(1 + q^2)^2} & 0 & \frac{\sqrt{2} q \sqrt{1 + q^2 + q^4}}{(1 + q^2)^2} \\
  	0 & 0 & 1 & 0 \\
  	\frac{1 + q^2 + q^4}{(1 + q^2)^2} & -\frac{(\sqrt{2} q \sqrt{1 + q^2 + q^4}}{(1 + q^2)^2} & 0 &	\frac{q^2}{(1 + q^2)^2} \\
  \end{pmatrix}.
  \label{m43221}
\end{equation}
 Changing lines of the matrix $\mathcal{U'}_{[4,3,2,2,1]}$ we get blocks $3 \times 3$ and $1 \times 1$.

\subsection{[4,4,2,2]}

\begin{multline}
\mathcal{R}_{[4,4,2,2]} =  {\rm diag}
(-q^{\varkappa_{[4,2,1,1]}}, \
 -q^{\varkappa_{[3,3,2]}} , \
 q^{\varkappa_{[4,2,2]}}, \
 q^{\varkappa_{[4,3,1]}} , \
 -q^{\varkappa_{[4,3,1]}})
= \\ {\rm diag}(
-1, \
 -1, \
 q^{-2} , \
 q^{-4}, \
 -q^{-4} )
\end{multline}

$\mathcal{R}_{[4,4,2,2]}$ has a pair of accidentally coinciding eigenvalues. We rotate $\mathcal{R}_{[4,4,2,2]}$ in the sector $1-2$ with

\begin{equation}
\boxed{\begin{array}{ll}
	c_{12}^{[4,4,2,2]} = \frac{1}{\sqrt{2}}, & s_{12}^{[4,4,2,2]} = -\frac{1}{\sqrt{2}}.
	\end{array}}
\end{equation}

and get

\begin{equation}
\mathcal{U'}_{[4,4,2,2]} =
\begin{pmatrix}
x_{11} & 0 & 0 & x_{14} & 0 \\
0 & x_{22} & x_{23} & 0 & -x_{25} \\
0 & x_{23} & x_{33} & 0 & x_{23} \\
-x_{14} & 0 & 0 & x_{11} & 0 \\
0 & x_{25} & x_{23} & 0 & -x_{22} \\
\end{pmatrix},
\end{equation}
where
\begin{equation}
\begin{array}{lll}
x_{11} = \frac{q^2}{1 + q^4}, &
x_{22} = -\frac{q^2}{(1 + q^2)^2} , &

x_{23} = -\frac{\sqrt{2} q \sqrt{1 - q + q^2} \sqrt{1 + q + q^2}}{(1 + q^2)^2} , \\
x_{25} = \frac{1 + q^2 + q^4}{(1 + q^2)^2}, &

x_{14} = \frac{\sqrt{1 - q + q^2}\sqrt{1 + q + q^2} \sqrt{1 - q^2 + q^4}}{
1 + q^4},&
x_{33} = -\frac{1+q^4}{(1+q^2)^2}. \\
\end{array}
\end{equation}

Changing lines in the matrix we get blocks $3 \times 3 $ and $2 \times 2$.

\subsection{[4,3,3,1,1]}

\begin{multline}
\mathcal{R}_{[4,3,3,1,1]} =  {\rm diag}(
q^{\varkappa_{[3,3,1,1]}} , \
 q^{\varkappa_{[3,3,1,1]}}, \
 -q^{\varkappa_{[4,2,1,1]}}, \
 -q^{\varkappa_{[3,3,2]}} , \
q^{\varkappa_{[4,3,1]}} , \
 -q^{\varkappa_{[4,3,1]}} )
= \\  {\rm diag}(
q^{2}, \
 q^{2} , \
 -1, \
 -1, \
 q^{-4}, \
 -q^{-4} )
\end{multline}
 $\mathcal{R}_{[4,3,3,1,1]}$ has two pairs of coinciding eigenvalues. We use rotation matrices in sectors $1-2$ and $3-4$ with:
\begin{equation}
\begin{array}{l}
c_{12}^{[4,3,3,1,1]} = \frac{1}{\sqrt{2}}(1+q^4) \sqrt{\frac{1+q^2+q^4}{1+2q^4+2q^8+q^12}}, \\
s_{12}^{[4,3,3,1,1]} = \frac{1}{\sqrt{2}} \sqrt{\frac{1 - q^2 + q^4 + q^8 - q^10 + q^12}{1 + 2 q^4 + q^6 + 2 q^8 + q^12} },
\end{array}
\end{equation}
and
\begin{equation}
\boxed{\begin{array}{ll}
	c_{34}^{[4,3,3,1,1]} = \frac{1}{\sqrt{2}}, & s_{34}^{[4,3,3,1,1]} = \frac{1}{\sqrt{2}}.
	\end{array}}
\end{equation}
and get

\begin{multline}
\mathcal{U'}_{[4,3,3,1,1]} =\\
\begin{pmatrix}
-\frac{q^2 + q^6}{1 + q^2 + q^6 + q^8} & 0 &
		0 & A_{14} & 0 &
	  A_{16} \\	
	0 & \frac{q^2}{1 + q^2 + q^4} & -\frac{q}{\sqrt{1 + q^2 + q^4}} & 0 & \frac{\sqrt{
		1 + q^2 + q^4 + q^6 + q^8}}{1 + q^2 + q^4} & 0 \\
		0 & \frac{q}{\sqrt{1 + q^2 + q^4}}  & \frac{-1 + q^2 - q^4}{1 + q^4} &
		0 & -q \frac{\sqrt{\frac{(1 + q^2 + q^4 + q^6 + q^8)}{1 + q^2 + q^4}}}{
		1 + q^4} & 0 \\
A_{41} & 0 &
		0 ,& \frac{1 + q^2 + q^4}{(1 + q^2)^2} & 0 &
		q \frac{\frac{\sqrt{1 + q^2 + q^4 + q^6 + q^8}}{(1 - q^2 + q^4)}}{(1 + q^2)^2} \\
			0 & \frac{\sqrt{1 + q^2 + q^4 + q^6 + q^8}}{1 + q^2 + q^4} &
		 \frac{ q \sqrt{\frac{1 + q^2 + q^4 + q^6 + q^8}{1 + q^2 + q^4}}}{1 + q^4} & 0 &
		\frac{q^4}{1 + q^2 + 2 q^4 + q^6 + q^8} & 0 \\
		A_{61} & 0 & 0 &
		 \frac{q \sqrt{ \frac{1 + q^2 + q^4 + q^6 + q^8}{1 - q^2 + q^4}  } }{(1 + q^2)^2} &
		0 & -\frac{q^4}{1 + q^2 + q^6 + q^8} \\
\end{pmatrix}
\label{rot43311}
\end{multline}

where
\begin{equation}
\begin{array}{ll}
 A_{41} = 	\frac{q + q^9}{(1 + q^2)^2 \sqrt{1 - q^2 + q^4 + q^8 - q^{10} + q^{12}}} &

A_{61} = 	-\frac{\sqrt{(1 + q^8) (1 + q^2 + q^4 + q^6 + q^8)}}{1 + q^2 + q^6 + q^8}
\\
A_{14} = -\frac{q (1 + q^8)}{(1 + q^2)^2 \sqrt{
		1 - q^2 + q^4 + q^8 - q^{10} + q^{12}}} &
	
A_{16} = 	\frac{\sqrt{(1 + q^8) (1 + q^2 + q^4 + q^6 + q^8)}}{1 + q^2 + q^6 + q^8}
	
\end{array}	
\end{equation}
Changing lines of the matrix $\mathcal{U'}_{[4,3,3,1,1]}$ we can see that it has a block-diagonal structure with two blocks $3 \times 3$

\subsection{[4,4,3,1]}
\begin{multline}
\mathcal{R}_{[4,4,3,1]} =  {\rm diag}(
q^{\varkappa_{[3,3,1,1]}} , \
 -q^{\varkappa_{[4,2,1,1]}}, \
 -q^{\varkappa_{[3,3,2]}} , \
 q^{\varkappa_{[4,2,2]}} , \
 q^{\varkappa_{[4,3,1]}} , \
 q^{\varkappa_{[4,3,1]}} , \
 -q^{\varkappa_{[4,3,1]}} , \
 -q^{\varkappa_{[4,3,1]}} , \
 q^{\varkappa_{[4,4]}} )
= \\
 {\rm diag}(
q^{2} , \
 -1 , \
 -1, \
 q^{-2} , \
 q^{-4} , \
q^{-4} , \
 -q^{-4} ,\
 -q^{-4} ,\
 q^{-8} )
\end{multline}

$\mathcal{R}_{[4,4,3,1]}$ has three pairs of coinciding eigenvalues: $\lambda_2 = \lambda_3$ (accidentally), $\lambda_5 = \lambda_6$, $\lambda_7 = \lambda_8$. We rotate corresponding Racah matrix $\mathcal{U}^{[4,4,3,1]}$ just in sectors 2-3 and 7-8, because $U^{[4,4,3,1]}_{56} = U^{[4,4,3,1]}_{65} = 0$.

To get just one rotation matrix in the sector $2-3$ we multiply the third line of $\mathcal{U^{[4,4,3,1]}}$ with $-1$. And we have
\begin{equation}
\boxed{\begin{array}{ll}
	c_{23}^{[4,4,3,1]} = \frac{1}{\sqrt{2}}, & s_{23}^{[4,4,3,1]} = -\frac{1}{\sqrt{2}}.
	\end{array}}
\end{equation}

Second rotation is in the sector $7-8$. Rotation matrices are with
\begin{equation}
\begin{array}{ll}
c_{1,78}^{[4,4,3,1]} = \sqrt{\frac{1+q^2+q^4}{2(1+q^4)}},  & s_{1,78}^{[4,4,3,1]} = - \sqrt{\frac{1-q^2+q^4}{2(1+q^4)}} \\
\end{array}
\end{equation}
and \begin{equation}
\begin{array}{ll}
	c_{2, 78}^{[4,4,3,1]} = 1$ , $ s_{2,78}^{[4,4,3,1]} = 0 .
\end{array}
\end{equation}

Finally we get:
\begin{equation}
\left(
\begin{array}{ccccccccc}
\text{x11} & 0 & \text{x13} & \text{x14} & 0 & \text{x16} & \text{x17} & 0 & \text{x19} \\
0 & 0 & 0 & 0 & \frac{1}{\sqrt{2}} & 0 & 0 & -\frac{1}{\sqrt{2}} & 0 \\
-\text{x13} & 0 & -\frac{2 q^2}{\left(q^2+1\right)^2} & -\text{x34} & 0 & -\frac{1}{\sqrt{\frac{4 q^2}{q^4-q^2+1}+2}} & \text{x37} & 0 & \text{x39} \\
\text{x14} & 0 & \text{x34} & \frac{q^8+q^6+3 q^4+q^2+1}{\left(q^2+1\right)^2 \left(q^4+q^2+1\right)} & 0 & -\frac{q}{\sqrt{q^4+q^2+1}} & -\frac{q
	\left(q^4+1\right)}{\left(q^2+1\right)^2 \sqrt{q^4+q^2+1}} & 0 & \text{x49} \\
0 & -\frac{1}{\sqrt{2}} & 0 & 0 & -\frac{1}{2} & 0 & 0 & -\frac{1}{2} & 0 \\
-\text{x16} & 0 & -\frac{1}{\sqrt{\frac{4 q^2}{q^4-q^2+1}+2}} & \frac{q}{\sqrt{q^4+q^2+1}} & 0 & \frac{-q^4+q^2-1}{2 \left(q^4+q^2+1\right)} &
\frac{1}{2} & 0 & -\text{x69} \\
-\text{x17} & 0 & \text{x27} & \frac{q^5+q}{\left(q^2+1\right)^2 \sqrt{q^4+q^2+1}} & 0 & \frac{1}{2} & -\frac{q^8+3 q^6-4 q^4+3 q^2+1}{2
	\left(q^8+q^6+q^2+1\right)} & 0 & \text{x79} \\
0 & -\frac{1}{\sqrt{2}} & 0 & 0 & \frac{1}{2} & 0 & 0 & \frac{1}{2} & 0 \\
\text{x19} & 0 & -\text{x39} & \text{x49} & 0 & \text{x69} & -\text{x79} & 0 & \text{x99} \\
\end{array}
\right),
\label{m4431}
\end{equation}
where
\begin{equation}
\begin{array}{ll}
x_{11} = \frac{q^4 \left(\left(q^2+1\right) \left(q^8+2 q^4+q^2+1\right) q^2+1\right)}{\left(q^2+1\right)^2 \left(\left(q^4+1\right) \left(q^{10}+q^8+q^6+q^4+2
	q^2+1\right) q^2+1\right)} &
x_{13} = \frac{\sqrt{2} q^3 \sqrt{\frac{q^{12}+q^{10}+q^8+q^6+q^4+q^2+1}{\left(q^4+1\right) \left(q^{10}+q^8+q^6+q^4+2 q^2+1\right) q^2+1}}}{\left(q^2+1\right)^2} \\
x_{14} = \frac{q^2 \sqrt{\frac{q^{12}+q^{10}+q^8+q^6+q^4+q^2+1}{\left(q^2+1\right) \left(q^8+q^6+2 q^4+q^2+2\right) q^2+1}}}{\left(q^2+1\right)^2} &
x_{16} = \frac{q \sqrt{\frac{q^{12}+q^{10}+q^8+q^6+q^4+q^2+1}{q^8+q^6+q^4+q^2+1}}}{q^4+q^2+1}\\
x_{17} = \frac{\left(q^5+q\right) \sqrt{\frac{q^{12}+q^{10}+q^8+q^6+q^4+q^2+1}{q^8+q^6+q^4+q^2+1}}}{q^8+q^6+q^2+1} &
x_{19} = \frac{\left(q^4+1\right) \sqrt{\frac{q^{12}+q^{10}+q^8+q^6+q^4+q^2+1}{q^4+q^2+1}}}{q^8+q^6+q^4+q^2+1} \\
x_{34} = \frac{\sqrt{2} q \sqrt{q^4-q^2+1}}{\left(q^2+1\right)^2} &

x_{37} = -\frac{\left(q^2-1\right)^2 \sqrt{\frac{2 q^2}{q^4-q^2+1}+1}}{\sqrt{2} \left(q^2+1\right)^2} \\
x_{39} = \frac{\left(q^4+1\right) \sqrt{\frac{q^{12}+q^{10}+q^8+q^6+q^4+q^2+1}{q^4+q^2+1}}}{q^8+q^6+q^4+q^2+1} &
x_{49} = \frac{q^6+q^2}{\left(q^4+q^2+1\right) \sqrt{q^8+q^6+q^4+q^2+1}} \\
x_{69} = \frac{q^3}{\sqrt{\left(q^2+1\right) \left(q^8+q^6+2 q^4+q^2+2\right) q^2+1}} &
x_{79} = \frac{q^3}{\sqrt{\left(q^4+q^2+1\right) \left(q^8+q^6+q^4+q^2+1\right)}} \\
x_{99}  = \frac{q^6}{\left(q^2+1\right) \left(q^8+q^6+2 q^4+q^2+2\right) q^2+1} & \\
\end{array}
\end{equation}
Changing lines of the matrix (\ref{m4431}) we can see that it has a block-diagonal structure with blocks $3 \times 3$ and $6 \times 6$.

\section{Representation [3,2] \label{s32}}
\begin{equation}
\begin{array}{l}
[3,2]^{\otimes 2} = [6,4] \oplus \underline{[6,3,1]} \oplus [6,2,2] \oplus \underline{[5,5]} \oplus [5,4,1] \oplus \underline{[5,4,1]} \oplus [5,3,2] \oplus \underline{[5,3,2]} \oplus [5,3,1,1]\oplus \\ \\
 \underline{[5,2,2,1]} \oplus ,[4,4,2] \oplus \underline{[4,4,1,1]} \oplus \underline{[4,3,3]} \oplus [4,3,2,1] \oplus \underline{[4,3,2,1]} \oplus [4,2,2,2] \oplus [3,3,3,1] \oplus \underline{[3,3,2,2]},
\end{array}
\end{equation}
where antisymmetric irreducible representations are underlined. Corresponding $\mathcal{R}-$matrices have eigenvalues:

\begin{equation}
\begin{array}{lll}
\lambda_{[6,4]} = q^{-17}, &
\lambda_{[6,3,1]} =- q^{-13}, &
\lambda_{[6,2,2]} = q^{-11}, \\
\lambda_{[5,5]} = -q^{-15}, &
\lambda_{[5,4,1]_{\pm}} = \pm q^{-10}, &
\lambda_{[5,3,2]_{\pm}} = \pm q^{-7}, \\
\lambda_{[5,3,1,1]} = q^{-5}, &
\lambda_{[5,2,2,1]} = -q^{-3}, &
\lambda_{[4,4,2]} = q^{-5}, \\
\lambda_{[4,4,1,1]} = -q^{-3}, &
\lambda_{[4,3,3]} = -q^{-3}, &
\lambda_{[4,3,2,1]_{\pm}} = \pm 1,  \\
\lambda_{[4,2,2,2]} = q^{3}, &
\lambda_{[3,3,3,1]} = q^{3},&
\lambda_{[3,3,2,2]} = -q^{5}.  \\
\end{array}
\end{equation}
One can see that several eigenvalues accidentally coincide:
\begin{equation}
\begin{array}{l}
 \lambda_{[5,3,1,1]} = \lambda_{[4,4,2]} =  q^{-5} \\
\lambda_{[5,2,2,1]} = \lambda_{[4,4,1,1]} =  \lambda_{[4,3,3]} = -q^{-3} \\
\lambda_{[4,2,2,2]} = \lambda_{[3,3,3,1]} = q^{3}. \\
\end{array}
\end{equation}

\begin{table}[h]
	\begin{center}
		\begin{tabular}{|c|c|c|c|c|c|}
			\hline
		\multicolumn{6}{|c|}{\rule{0cm}{0.2cm}} \\
		\multicolumn{6}{|c|}{Table of matrices with accidentally coinciding eigenvalues} \\
		\multicolumn{6}{|c|}{$ [3,2]$} \\ [0.2cm]
		
		\hline
		
		size & repr &   \multicolumn{2}{|c|}{eigenvalues} & result  & blocks \\
		\cline{3-4}
		& & repetitive & accidental & & \\
		
			\hline
			& & & & &\\
			\rule{0cm}{0.5cm}
			6 & [4,3,3,2,2,1]	&$ \lambda_5 = \lambda_6 $ & $ \lambda_3 = \lambda_4 $ & + & (3,3)\\ [0.2cm]
			& [6,4,2,1,1,1] &$ \lambda_1 = \lambda_2 $ & $ \lambda_3 = \lambda_4 $ & + & (3,3)\\ [0.2cm]
	
			\hline
	%
			\end{tabular}
	\end{center}
\end{table}

\subsection{[4,3,3,2,2,1]}
\begin{multline}
\mathcal{R}_{[4,3,3,2,2,1]} = {\rm diag}(
-q^{\varkappa_{[4,3,2,1]}} , \
q^{\varkappa_{[4,3,2,1]}} , \
 q^{\varkappa_{[4,2,2,2]}}  , \
 q^{\varkappa_{[3,3,3,1]}} , \
-q^{\varkappa_{[3,3,2,2]}} , \
 - q^{\varkappa_{[3,3,2,2]}} )
= \\ {\rm diag}(
-1 , \
1 , \
 q^3 , \
 q^3 , \
 -q^5 , \
 -q^5 ).
\end{multline}

$\mathcal{R}_{[4,3,3,2,2,1]}$ has two pairs of coinciding eigenvalues: $\lambda_3 = \lambda_4$ (accidentally) and $\lambda_5 = \lambda_6$.
To get just one rotation matrix in  the sector $3 - 4$ we change the sign of the third line of the  Racah matrix $\mathcal{U}_{[4,3,3,2,2,1]}$. For the $\mathcal{O}^{[4,3,3,2,2,1]}_{34}$ we use

\begin{equation}
\boxed{\begin{array}{ll}
	c_{34}^{[4,3,3,2,2,1]} = \frac{1}{\sqrt{2}}, & s_{34}^{[4,3,3,2,2,1]} = -\frac{1}{\sqrt{2}}.
	\end{array}}
\end{equation}
For the second rotation in the sector $5-6$ we use two rotation  matrices with

\begin{equation}
\begin{array}{ll}
c_{1,56}^{[4,3,3,2,2,1]} = \frac{\sqrt{\frac{q \left(q^2+1\right) \left(q^8+1\right)}{q^{12}+2 q^{10}+2 q^8+q^6+2 q^4+2 q^2+1}+1}}{\sqrt{2}}, &
s_{1,56}^{[4,3,3,2,2,1]} = \frac{\sqrt{1-\frac{q \left(q^2+1\right) \left(q^8+1\right)}{q^{12}+2 q^{10}+2 q^8+q^6+2 q^4+2 q^2+1}}}{\sqrt{2}} \\

c_{2,56}^{[4,3,3,2,2,1]} = \frac{\sqrt{1-\frac{q \left(q^2+1\right) \left(q^8+1\right)}{q^{12}+2 q^{10}+2 q^8+q^6+2 q^4+2 q^2+1}}}{\sqrt{2}} &
s_{2,56}^{[4,3,3,2,2,1]} = \frac{\sqrt{\frac{q \left(q^2+1\right) \left(q^8+1\right)}{q^{12}+2 q^{10}+2 q^8+q^6+2 q^4+2 q^2+1}+1}}{\sqrt{2}}
\end{array}
\end{equation}
Finally, we get a matrix:

\begin{multline}
\mathcal{U'}_{[4,3,3,2,2,1]} =  \\
\left(
\begin{array}{cccccc}
\frac{q^3}{q^6+q^4-q^3+q^2+1} & 0 & 0 & x_{14} & -x_{15} & 0 \\
0 & -\frac{q^3}{q^6+q^4+q^3+q^2+1} & x_{23} & 0 & 0 & x_{26} \\
0 & -x_{23} & -x_{33} & 0 & 0 & -x_{36} \\
x_{14} & 0 & 0 & x_{44} & x_{45} & 0 \\
x_{15} & 0 & 0 & -x_{45} & \frac{q^2 ((q-1) q+1)}{\left(q^2+1\right) \left((q-1) q \left(q^2+1\right)+1\right)} & 0 \\
0 & -x_{26} & -x_{36} & 0 & 0 & \frac{q^2 \left(q^2+q+1\right)}{\left(q^2+1\right) \left(q^4+q^3+q^2+q+1\right)} \\
\end{array}
\right),
\end{multline}
where
\begin{equation}
\begin{array}{ll}
x_{14 } = \frac{q \sqrt{\left(q^4+1\right) \left((q-1) q \left(q^2+1\right)+1\right)}}{q^6+q^4-q^3+q^2+1}, &
x_{15}  =\frac{\sqrt{\left(q^4+1\right) \left(\left(q^2+1\right) \left(q^4-q+1\right) q^2+1\right)}}{q^6+q^4-q^3+q^2+1} ,\\
x_{23} = \frac{q \sqrt{\left(q^4+1\right) \left(q^4+q^3+q^2+q+1\right)}}{q^6+q^4+q^3+q^2+1} ,&
x_{26} = \frac{\sqrt{\left(q^4+1\right) \left(\left(q^2+1\right) \left(q^4+q+1\right) q^2+1\right)}}{q^6+q^4+q^3+q^2+1} ,\\
x_{33} = \frac{q^2}{((q-1) q+1) \left(q^2+1\right)}-1, &
x_{36} = \frac{q \sqrt{\left(q^4+q^3+q^2+q+1\right) \left(\left(q^2+1\right) \left(q^4+q+1\right) q^2+1\right)}}{\left(q^6+2 q^4+q^3+2 q^2+q+2\right) q^2+1} ,\\
x_{44} = \frac{q^2}{q^4+q^3+2 q^2+q+1}-1 ,&
x_{45} = \frac{q \left(-(q-1) q ((q-1) q+1) \left(q^2+q+1\right)-1\right)}{\left(q^2+1\right) \sqrt{\left((q-1) q \left(q^2+1\right)+1\right)
		\left(\left(q^2+1\right) \left(q^4-q+1\right) q^2+1\right)}}. \\
\end{array}
\end{equation}
One can see that there are blocks $3 \times 3$ and $3 \times 3$ in $\mathcal{U'}_{[4,3,3,2,2,1]}$.

\subsection{[6,4,2,1,1,1]}
\begin{multline}
\mathcal{R}_{[6,4,2,1,1,1]} =  {\rm diag}(
q^{\varkappa_{[5,3,1,1]}} ,\
q^{\varkappa_{[5,3,1,1]}} , \
 -q^{\varkappa_{[5,2,2,1]}} , \
- q^{\varkappa_{[4,4,1,1]}} , \
 -q^{\varkappa_{[4,3,2,1]}} , \
 q^{\varkappa_{[4,3,2,1]}} )
= \\  {\rm diag}(
q^{-5}, \
 q^{-5} , \
 -q^{-3} , \
-q^{-3} , \
 -1, \
 1 ).
\end{multline}

$\mathcal{R}_{[6,4,2,1,1,1]}$ has two pairs of coinciding eigenvalues: $\lambda_1 = \lambda_2 $ and $\lambda_3 = \lambda_4$ (accidentally).
The rotation in the sector $3 - 4$ should be done with
\begin{equation}
\boxed{\begin{array}{ll}
	c_{34}^{[6,4,2,1,1,1]} = \frac{1}{\sqrt{2}}, & s_{34}^{[6,4,2,1,1,1]} = -\frac{1}{\sqrt{2}}.
	\end{array}}
\end{equation}
and rotation in the sector $1-2$ requires two rotation matrices with
\begin{multline}
c_{1,12}^{[6,4,2,1,1,1]} = \frac{ \sqrt{\frac{\left(q^{14}+q^{10}+q^8+q^7+q^6+q^4+1\right)^2((q-1) q+1) \left(q^6+q^5+q^4+q^3+q^2+q+1\right) \left((q-1) q
			\left(q^2+1\right)+1\right)}{\left(q^4+1\right) \left(q^{34}+2 q^{32}+3 q^{30}+5 q^{28}+8 q^{26}+11 q^{24}+14 q^{22}+17 q^{20}+19 q^{18}+20 q^{16}+18
			q^{14}+17 q^{12}+15 q^{10}+11 q^8+7 q^6+5 q^4+4 q^2+2\right) q^2+1}}}{\sqrt{2}} ,\\
s_{1,12}^{[6,4,2,1,1,1]} = \frac{ \sqrt{\frac{q \left(q^{14}+q^{10}+q^8-q^7+q^6+q^4+1\right)^2\left(q^2+1\right) \left(q^9+q^8+q^7+q^5+2 q+1\right)+1}{\left(q^4+1\right) \left(q^{34}+2
			q^{32}+3 q^{30}+5 q^{28}+8 q^{26}+11 q^{24}+14 q^{22}+17 q^{20}+19 q^{18}+20 q^{16}+18 q^{14}+17 q^{12}+15 q^{10}+11 q^8+7 q^6+5 q^4+4 q^2+2\right)
			q^2+1}}}{\sqrt{2}}   ,\\
c_{2,12}^{[6,4,2,1,1,1]} = \frac{\sqrt{\frac{q \left(q^2+1\right) \left(q^8+1\right)}{q^{12}+2 q^{10}+2 q^8+q^6+2 q^4+2 q^2+1}+1}}{\sqrt{2}}  ,\\
s_{2,12}^{[6,4,2,1,1,1]} = -\frac{\sqrt{1-\frac{q \left(q^2+1\right) \left(q^8+1\right)}{q^{12}+2 q^{10}+2 q^8+q^6+2 q^4+2 q^2+1}}}{\sqrt{2}}   .\\
\end{multline}

\begin{multline}
\mathcal{U'}_{[6,4,2,1,1,1]} = \\
\left(
\begin{array}{cccccc}
\frac{q^2 \left(-q^2+q-1\right)}{\left(q^2+1\right) \left((q-1) q \left(q^2+1\right)+1\right)} & 0 & 0 & x_{14} & 0 & -x_{16} \\
0 & -\frac{q^2 \left(q^2+q+1\right)}{\left(q^2+1\right) \left(q^4+q^3+q^2+q+1\right)} & x_{23} & 0 & -x_{25} & 0 \\
0 & x_{23} & x_{33} & 0 & x_{35} & 0 \\
x_{14}& 0 & 0 & x_{44} & 0 &x_{46} \\
0 & x_{25} & -x_{35} & 0 & \frac{q^3}{q^6+q^4+q^3+q^2+1} & 0 \\
x_{16} & 0 & 0 & x_{46} & 0 & -\frac{q^3}{q^6+q^4-q^3+q^2+1} \\
\end{array}
\right),
\end{multline}

where

\begin{equation}
\begin{array}{ll}
x_{14} = \frac{q \left(-(q-1) q ((q-1) q+1) \left(q^2+q+1\right)-1\right)}{\left(q^2+1\right) \sqrt{\left((q-1) q \left(q^2+1\right)+1\right)
		\left(\left(q^2+1\right) \left(q^4-q+1\right) q^2+1\right)}},&
x_{16} = \frac{\sqrt{\left(q^4+1\right) \left(\left(q^2+1\right) \left(q^4-q+1\right) q^2+1\right)}}{q^6+q^4-q^3+q^2+1}, \\
x_{23} = -\frac{q \left(q^6+q^5+q^4+q^3+q^2+q+1\right)}{\left(q^2+1\right) \sqrt{\left(q^4+q^3+q^2+q+1\right) \left(\left(q^2+1\right) \left(q^4+q+1\right)
		q^2+1\right)}} , &
x_{25} = \frac{\sqrt{\left(q^4+1\right) \left(\left(q^2+1\right) \left(q^4+q+1\right) q^2+1\right)}}{q^6+q^4+q^3+q^2+1}, \\
x_{33} = \frac{q^2}{((q-1) q+1) \left(q^2+1\right)}-1 , &
x_{35} = \frac{q \sqrt{\left(q^4+1\right) \left(q^4+q^3+q^2+q+1\right)}}{q^6+q^4+q^3+q^2+1} , \\
x_{44} = 1-\frac{q^2}{q^4+q^3+2 q^2+q+1}, &
x_{46} = -\frac{q \sqrt{\left(q^4+1\right) \left((q-1) q \left(q^2+1\right)+1\right)}}{q^6+q^4-q^3+q^2+1}.\\
\end{array}
\end{equation}
One can see that $\mathcal{U'}_{[6,4,2,1,1,1]}$ has a block-diagonal form with blocks $3 \times 3$ and $3 \times 3$.

\section{Representation [4,1] \label{s41}}
\begin{equation}
\begin{array}{l}
[4,1]^{\otimes 2} = [8, 2]\oplus\underline{[8, 1, 1]}\oplus\underline{[7, 3]}\oplus\underline{[7, 2, 1]}\oplus[7, 2, 1]\oplus[7, 1, 1, 1]\oplus[6, 4]\oplus\underline{[6, 3, 1]}\oplus[6, 3, 1]\oplus[6, 2, 2] \oplus \\ \\
\underline{[6, 2, 1, 1]}\oplus \underline{[5, 5]}\oplus\underline{[5, 4, 1]}\oplus [5, 4, 1] \oplus \underline{[5, 3, 2]}\oplus [5, 3, 1, 1]\oplus[4, 4, 2]\oplus\underline{[4, 4, 1, 1]}.
\end{array}
\end{equation}
Eigenvalues are:
\begin{equation}
\begin{array}{lll}
\lambda_{[8,2]} = q^{-27}, & \lambda_{[8,1,1]} = -q^{-25}, & \lambda_{[7,3]} = -q^{-21}, \\
\lambda_{[7,2,1]_{\pm}} = \pm q^{-18}, & \lambda_{[7,1,1,1]} = q^{-15}, & \lambda_{[6,4]} = q^{-17}, \\
 \lambda_{[6,3,1]_{\pm}} = \pm q^{-13},  & \lambda_{[6,2,2]} = q^{-11}, &
 \lambda_{[6,2,1,1]} = -q^{-9}, \\
 \lambda_{[5,5]} = -q^{-15}, & \lambda_{[5,4,1]_{\pm }} = \pm q^{-10}, &
 \lambda_{[5,3,2]} = -q^{-7}, \\
  \lambda_{[5,3,1,1]} = q^{-5}, & \lambda_{[4,4,2]} = q^{-5}, & \lambda_{[4,4,1,1]} = -q^{-3}, \\ \\
\end{array}
\end{equation}
where there is just one pair of accidentally coinciding eigenvalues:
\begin{equation}
  \lambda_{[5,3,1,1]} = \lambda_{[4,4,2]} = q^{-5}.
\end{equation}

\begin{table}[h]
	\begin{center}
		\begin{tabular}{|c|c|c|c|c|c|}
			\hline
		\multicolumn{6}{|c|}{\rule{0cm}{0.2cm}} \\
		\multicolumn{6}{|c|}{Table of matrices with accidentally coinciding eigenvalues} \\
		\multicolumn{6}{|c|}{$ [4,1]$} \\ [0.2cm]
		
		\hline
		
		size & repr &   \multicolumn{2}{|c|}{eigenvalues} & result  & blocks \\
		\cline{3-4}
		& & repetitive & accidental & & \\
		
			\hline
			& & & & & \\
			\rule{0cm}{0.5cm}
			4 & [5,4,3,2,1]	& - & $ \lambda_2 = \lambda_3 $ & + & (3,1)\\ [0.2cm]

			\hline
			& & & & &\\
			\rule{0cm}{0.5cm}
			6 & [5,5,3,2] &  $\lambda_3 = \lambda_4 $ &$ \lambda_5 = \lambda_6 $ & + & (3,3) \\ [0.2cm]
			&[5,4,4,1,1] & $\lambda_2 = \lambda_3 $, $ \lambda_4 = \lambda_5 $  & $ \lambda_6 = \lambda_7 $ & + & (3,3)\\ [0.2cm]
			
			& [6,4,2,2,1]	& - & $ \lambda_4 = \lambda_5 $ & - &\\
[0.2cm]			
			\hline

		\end{tabular}
	\end{center}
\end{table}

\subsection{[5,4,3,2,1]}
\begin{multline}
\mathcal{R}_{[5,4,3,2,1]} =
 {\rm diag}(
-q^{\varkappa_{[5,3,2]}} , \
 q^{\varkappa_{[5,3,1,1]}}, \
 q^{\varkappa_{[4,4,2]}}, \
 -q^{\varkappa_{[4,4,1,1]}} )
=
 {\rm diag}(
-q^{-7}, \
 q^{-5} , \
 q^{-5} , \
 -q^{-3} ).
\end{multline}
There are just one pair of coinciding eigenvalues: $\lambda_2 = \lambda_3$. The rotation matrix is with

\begin{equation}
\boxed{\begin{array}{ll}
	c_{23}^{[5,4,3,2,1]} = \frac{1}{\sqrt{2}}, & s_{23}^{[5,4,3,2,1]} = -\frac{1}{\sqrt{2}}.
	\end{array}}
\end{equation}

After the rotation $\mathcal{U}_{[5,4,3,2,1]}$ goes to
\begin{equation}
\mathcal{U'}_{[5,4,3,2,1]} =
\left(
\begin{array}{cccc}
\frac{q^2}{\left(q^2+1\right)^2} & \frac{\sqrt{2} q \sqrt{q^4+q^2+1}}{\left(q^2+1\right)^2} & 0 & \frac{q^4+q^2+1}{\left(q^2+1\right)^2} \\
\frac{\sqrt{2} q \sqrt{q^4+q^2+1}}{\left(q^2+1\right)^2} & \frac{q^4+1}{\left(q^2+1\right)^2} & 0 & -\frac{\sqrt{2} q
	\sqrt{q^4+q^2+1}}{\left(q^2+1\right)^2} \\
0 & 0 & -1 & 0 \\
\frac{q^4+q^2+1}{\left(q^2+1\right)^2} & -\frac{\sqrt{2} q \sqrt{q^4+q^2+1}}{\left(q^2+1\right)^2} & 0 & \frac{q^2}{\left(q^2+1\right)^2} \\
\end{array}
\right),
\end{equation}
where one can see $3 \times 3$  and $1 \times 1$ blocks.

\subsection{[5,5,3,2]}
\begin{multline}
\mathcal{R}_{[5,5,3,2]} =
 {\rm diag}(
-q^{\varkappa_{[5,4,1]}} , \
 q^{\varkappa_{[5,4,1]}} , \
 -q^{\varkappa_{[5,3,2]}}, \
- q^{\varkappa_{[5,3,2]}} , \
q^{\varkappa_{[5,3,1,1]}}, \
 q^{\varkappa_{[4,4,2]}} )
= \\
 {\rm diag}(
-q^{-10}, \
 q^{-10} , \
 -q^{-7}, \
-q^{-7} , \
 q^{-5}, \
 q^{-5} ).
\end{multline}
There are two pairs of coinciding eigenvalues: $\lambda_3 = \lambda_4 $ and $\lambda_5 = \lambda_6$ (accidentally).
${\rm cos}$ and ${\rm sin}$ for rotation in the sector $5-6$ are
\begin{equation}
\boxed{\begin{array}{ll}
	c_{56}^{[5,5,3,2]} = \frac{1}{\sqrt{2}}, & s_{56}^{[5,5,3,2]} = -\frac{1}{\sqrt{2}}.
	\end{array}}
\end{equation}

For the second rotation one needs two matrices with

\begin{equation}
\begin{array}{ll}
c_{1,34}^{[5,5,3,2]} = \frac{\sqrt{\frac{q \left(q^4+1\right) \left(q^2+1\right)^3}{((q-1) q+1) \left(q^2+1\right) \left(q^2+q+1\right) \left(q^4+2\right) q^2+1}+1}}{\sqrt{2}}, &
s_{1,34}^{[5,5,3,2]} = -\frac{\sqrt{1-\frac{q \left(q^2+1\right)^3 \left(q^4+1\right)}{((q-1) q+1) \left(q^2+1\right) \left(q^2+q+1\right) \left(q^4+2\right) q^2+1}}}{\sqrt{2}}, \\
c_{2,34}^{[5,5,3,2]} = \frac{\sqrt{\frac{q \left(q^2+1\right) \left(q^4+1\right)^2}{((q-1) q+1) \left(q^2+q+1\right) \left(q^6+q^4+2\right) q^2+1}+1}}{\sqrt{2}}, &
s_{2,34}^{[5,5,3,2]} = -\frac{\sqrt{\frac{((q-1) q+1) \left(q^4+q^3+q^2+q+1\right) \left((q-1) q ((q-1) q+1) \left(q^2+q+1\right)+1\right)}{((q-1) q+1) \left(q^2+q+1\right)
			\left(q^6+q^4+2\right) q^2+1}}}{\sqrt{2}}. \\
\end{array}
\end{equation}
After the rotations we get
\begin{multline}
\mathcal{U'}_{[5,5,3,2]} = \\
\left(
\begin{array}{cccccc}
\frac{q^3}{q^6+q^4+q^3+q^2+1} & 0 & x_{13} & 0 & x_{15} & 0 \\
0 & -\frac{q^3}{q^6+q^4-q^3+q^2+1} & 0 & x_{24} & 0 & -x_{26} \\
x_{13} & 0 & x_{33} & 0 & x_{35} & 0 \\
0 & x_{24} & 0 &x_{44} & 0 & -x_{46} \\
x_{15} & 0 & x_{35} & 0 & \frac{q^2 \left(q^2+q+1\right)}{\left(q^2+1\right) \left(q^4+q^3+q^2+q+1\right)} & 0 \\
0 & x_{26} & 0 & x_{46} & 0 & \frac{q^2 \left(-q^2+q-1\right)}{\left(q^2+1\right) \left((q-1) q \left(q^2+1\right)+1\right)} \\
\end{array}
\right),
\end{multline}
where
\begin{equation}
\begin{array}{ll}
x_{13} = \frac{q \sqrt{\left(q^4+1\right) \left(q^4+q^3+q^2+q+1\right)}}{q^6+q^4+q^3+q^2+1}, &
x_{15} = \frac{\sqrt{((q-1) q+1) \left(q^4+1\right) \left(q^6+q^5+q^4+q^3+q^2+q+1\right)}}{q^6+q^4+q^3+q^2+1}, \\
x_{24} = -\frac{q \sqrt{\left(q^4+1\right) \left((q-1) q \left(q^2+1\right)+1\right)}}{q^6+q^4-q^3+q^2+1},&
x_{26} = \frac{\sqrt{\left(q^2+q+1\right) \left(q^4+1\right) \left((q-1) q ((q-1) q+1) \left(q^2+q+1\right)+1\right)}}{q^6+q^4-q^3+q^2+1}, \\
x_{33} = q \left(\frac{1}{q^2+1}+\frac{1}{-q^2+q-1}\right)+1, &
x_{35} = -\frac{q \sqrt{((q-1) q+1) \left(q^4+q^3+q^2+q+1\right) \left(q^6+q^5+q^4+q^3+q^2+q+1\right)}}{\left(q^6+2 q^4+q^3+2 q^2+q+2\right) q^2+1}, \\
x_{44} = 1-\frac{q^2}{q^4+q^3+2 q^2+q+1}, &
x_{46} = q \sqrt{\frac{(q-1) q ((q-1) q+1) \left(q^2+q+1\right)+1}{\left(q^2+1\right) \left(q^6+q^4-q^3+q^2+1\right)}}. \\
\end{array}
\end{equation}
$\mathcal{U'}_{[5,5,3,2]} $ has blocks  $3 \times 3$ and $3 \times 3$.

\subsection{[5,4,4,1,1]}
\begin{multline}
\mathcal{R}_{[5,4,4,1,1]} =
 {\rm diag}(
-q^{\varkappa_{[5,4,1]}} , \
 q^{\varkappa_{[5,4,1]}} , \
 q^{\varkappa_{[5,3,1,1]}} , \
 q^{\varkappa_{[4,4,2]}} , \
 -q^{\varkappa_{[4,4,1,1]}} , \
 -q^{\varkappa_{[4,4,1,1]}} )
= \\
 {\rm diag}(
-q^{-10} , \
 q^{-10}, \
 q^{-5}, \
q^{-5}, \
 -q^{-3}, \
-q^{-3}).
\end{multline}
There are two pairs of coinciding eigenvalues: $\lambda_3 = \lambda_4 $ (accidentally)  and $\lambda_5 = \lambda_6$.
For the first rotation we use
\begin{equation}
\boxed{\begin{array}{ll}
	c_{34}^{[5,4,4,1,1]} = \frac{1}{\sqrt{2}}, & s_{34}^{[5,4,4,1,1]} = -\frac{1}{\sqrt{2}}.
	\end{array}}
\end{equation}
For the rotation in the sector $5-6$ we need two matrices with
\begin{equation}
\begin{array}{l}
c_{1,56}^{[5,4,4,1,1]} =\sqrt{\frac{q^{16}+q^{14}-q^{13}+q^{12}+2 q^{10}-q^9+q^8-q^7+2 q^6+q^4-q^3+q^2+1}{2(q^{16}+q^{14}+q^{12}+2 q^{10}+q^8+2
		q^6+q^4+q^2+1)}} ,\\
s_{1,56}^{[5,4,4,1,1]} =\sqrt{-\frac{\left(q^6+q^3+1\right) \left(q^{10}+q^8+q^6-q^5+q^4+q^2+1\right)}{2(-q^{26}-2 q^{24}-3 q^{22}+q^{21}-5 q^{20}+q^{19}-6
		q^{18}+q^{17}-8 q^{16}+2 q^{15}-8 q^{14}+q^{13}-8 q^{12}+2 q^{11}-8 q^{10}+q^9-6 q^8+q^7-5 q^6+q^5-3 q^4-2 q^2-1)}} ,\\
	c_{2,56}^{[5,4,4,1,1]} = \sqrt{\frac{q^{16}+q^{14}-q^{13}+q^{12}-2 q^{11}+2 q^{10}-q^9+3 q^8-q^7+2 q^6-2 q^5+q^4-q^3+q^2+1}{2(q^{16}+q^{14}+q^{12}+2 q^{10}+3
			q^8+2 q^6+q^4+q^2+1)}},\\
	s_{2,56}^{[5,4,4,1,1]} = \sqrt{\frac{q^{16}+q^{14}+q^{13}+q^{12}+2 q^{11}+2 q^{10}+q^9+3 q^8+q^7+2 q^6+2 q^5+q^4+q^3+q^2+1}{2(q^{16}+q^{14}+q^{12}+2 q^{10}+3
			q^8+2 q^6+q^4+q^2+1)}},\\
\end{array}
\end{equation}

\begin{equation}
\mathcal{U'}_{[5,4,4,1,1]} =
\begin{pmatrix}
x_{11} & 0 & x_{13} & 0 & 0 & x_{16} \\
0 & x_{22} & 0 & x_{24} & x_{25} & 0 \\
x_{13} & 0 & x_{33} & 0 & 0 & x_{36} \\
0 & -x_{24} & 0 & x_{44} & x_{45} & 0\\
0 & x_{25} & 0 & -x_{45} & x_{55} & 0 \\
x_{16} & 0 & x_{36} & 0 & 0 & x_{66} \\
\end{pmatrix}
\end{equation}
Elements of $\mathcal{U'}_{[5,4,4,1,1]}$ are pretty complicated. That's why we don't list them here.
$\mathcal{U'}_{[5,4,4,1,1]}$ has block-diagonal form with two blocks $3 \times 3$.

\section{Representation [4,2] \label{s42}}
\begin{equation}
\begin{array}{l}
[4,2]^{\otimes 2} = [8,4] \oplus \underline{[8,3,1]}\oplus [8,2,2] \oplus \underline{[7,5]} \oplus \underline{[7,4,1]} \oplus [7,4,1] \oplus \underline{[7,3,2]} \oplus[7,3,2] \oplus [7,3,1,1] \oplus \\ \\
\underline{[7,2,2,1]}\oplus [6,6] \oplus \underline{[6,5,1]} \oplus [6,5,1] \oplus \underline{[6,4,2]} \oplus2[6,4,2]\oplus \underline{6,4,1,1} \oplus \underline{[6,3,3]} \oplus {[6,3,2,1]} \oplus \\ \\
\underline{[6,3,2,1]} \oplus [6,2,2,2] \oplus \underline{[5,5,2]} \oplus [5,5,1,1] \oplus \underline{[5,4,3]} \oplus [5,4,3] \oplus \underline{[5,4,2,1]} \oplus [5,4,2,1] \oplus [5,3,3,1] \oplus \\ \\
\underline{[5,3,2,2]} \oplus [4,4,4] \oplus \underline{[4,4,3,1]} \oplus [4,4,2,2].
\end{array}
\end{equation}
Eigenvalues are:
\begin{equation}
\begin{array}{lllll}
\lambda_{[8,4]} = q^{-30}, &
\lambda_{[8,3,1]} =- q^{-26},&
\lambda_{[8,2,2]} = q^{-24}, &
\lambda_{[7,5]} = -q^{-26}, &
\lambda_{[7,4,1]_{\pm}} = \pm q^{-21},  \\
\lambda_{[7,3,2]_{\pm}} = \pm q^{-18}, &
\lambda_{[7,3,1,1]} = q^{-16},&
\lambda_{[7,2,2,1]} = -q^{-14}, &
\lambda_{[6,6]} = q^{-24}, &
\lambda_{[6,5,1]_{\pm}} = \pm q^{-18}, \\
\lambda_{[6,4,2]_{\pm}} = \pm q^{-14},&
\lambda_{[6,4,1,1]} = -q^{-12}, &
\lambda_{[6,3,3]} = -q^{-12},&
\lambda_{[6,3,2,1]_{\pm}} =\pm q^{-9},&
\lambda_{[6,2,2,2]} = q^{-6}, \\
\lambda_{[5,5,2]} = -q^{-12}, &
\lambda_{[5,5,1,1]} = q^{-10},&
\lambda_{[5,4,3]_{\pm}} = \pm q^{-9}, &
\lambda_{[5,4,2,1]_{\pm}} = \pm q^{-6}, &
\lambda_{[5,3,3,1]} = q^{-4}, \\
\lambda_{[5,3,3,2]} = -q^{-2}, &
\lambda_{[4,4,4]} = q^{-6}, &
\lambda_{[4,4,3,1]} = -q^{-2}, &
\lambda_{[4,4,2,2]} = 1. & \\
\end{array}
\end{equation}
Accidentally coinciding eigenvalues are:
\begin{equation}
\begin{array}{lll}
\lambda_{[8,3,1]} = \lambda_{[7,5]} = -q^{-26}, &
\lambda_{[8,2,2]} = \lambda_{[6,6]} = q^{-24}, &
\lambda_{[7,3,2]_{\pm}} = \lambda_{[6,5,1]_{\pm}} =\pm q^{-18}, \\
\lambda_{[7,2,2,1]} = \lambda_{[6,4,2]_{-}} = -q^{-14}, &
\lambda_{[6,4,1,1]} = \lambda_{[6,3,3]} = \lambda_{[5,5,2]} = -q^{-12}, &
\lambda_{[6,3,2,1]_{\pm}} = \lambda_{[5,4,3]_{\pm}} = \pm q^{-9}, \\
\lambda_{[6,2,2,2]} = \lambda_{[5,4,2,1]_{+}} = \lambda_{[4,4,4]}= q^{-6}, &
\lambda_{[5,3,2,2]} = \lambda_{[4,4,3,1]} = -q^{-2}. & \\
\end{array}
\end{equation}
\begin{table}[h]
	\begin{center}
		\begin{tabular}{|c|c|c|c|c|c|}
			\hline
		\multicolumn{6}{|c|}{\rule{0cm}{0.2cm}} \\
		\multicolumn{6}{|c|}{Table of matrices with accidentally coinciding eigenvalues} \\
		\multicolumn{6}{|c|}{$ [4,2]$} \\ [0.2cm]
		
		\hline
		
		size & repr &   \multicolumn{2}{|c|}{eigenvalues} & result  & blocks \\
		\cline{3-4}
		& & repetitive & accidental & & \\

			\hline
			& & & & & \\
			\rule{0cm}{0.5cm}
			4 & [5,4,3,3,2,1]	& - & $ \lambda_2 = \lambda_3 $ & + & (3,1)\\ [0.2cm]

			\hline
			& & & & & \\
			\rule{0cm}{0.5cm}
			5 & [5,5,3,3,1,1] & - & $\lambda_4 = \lambda_5 $ & + & (3,2) \\ [0.2cm]
            \hline
	
			& & & & & \\
			\rule{0cm}{0.5cm}
			6 & [5,4,4,2,2,1]	&  $\lambda_5 = \lambda_6 $ & $ \lambda_3 = \lambda_4 $ & + & (3,3)\\ [0.2cm]
			& [11,6,1]	&  $\lambda_1 = \lambda_2 $ & $ \lambda_3 = \lambda_4 $ & + & (3,3) \\ [0.2cm]
			\hline


		\end{tabular}
	\end{center}
\end{table}

\subsection{[5,4,3,3,2,1]}
\begin{multline}
\mathcal{R}_{[5,4,3,3,2,1]} =
{\rm diag}(
q^{\varkappa_{[5,3,3,1]}} , \
 -q^{\varkappa_{[5,3,2,2]}} , \
 -q^{\varkappa_{[5,5,3,1]}} , \
 q^{\varkappa_{[4,4,2,2]}} )
=
{\rm diag}(
q^{-4}, \
 -q^{-2} , \
 -q^{-2} ,\
 1).
\end{multline}
There is one pair of coinciding eigenvalues: $\lambda_2 = \lambda_3 $. We use a rotation matrix with
\begin{equation}
\boxed{\begin{array}{ll}
	c_{23}^{[5,4,3,3,2,1]} = \frac{1}{\sqrt{2}}, & s_{23}^{[5,4,3,3,2,1]} = -\frac{1}{\sqrt{2}}
	\end{array}}
\end{equation}
and get a block-diagonal Racah matrix:
\begin{equation}
\mathcal{U'}_{[5,4,3,3,2,1]} =
\left(
\begin{array}{cccc}
\frac{q^2}{\left(q^2+1\right)^2} & \frac{\sqrt{2} q \sqrt{q^4+q^2+1}}{\left(q^2+1\right)^2} & 0 & -\frac{q^4+q^2+1}{\left(q^2+1\right)^2} \\
-\frac{\sqrt{2} q \sqrt{q^4+q^2+1}}{\left(q^2+1\right)^2} & -\frac{q^4+1}{\left(q^2+1\right)^2} & 0 & -\frac{\sqrt{2} q
	\sqrt{q^4+q^2+1}}{\left(q^2+1\right)^2} \\
0 & 0 & 1 & 0 \\
-\frac{q^4+q^2+1}{\left(q^2+1\right)^2} & \frac{\sqrt{2} q \sqrt{q^4+q^2+1}}{\left(q^2+1\right)^2} & 0 & \frac{q^2}{\left(q^2+1\right)^2} \\
\end{array}
\right).
\end{equation}
Changing the order of lines in $\mathcal{U'}_{[5,4,3,3,2,1]}$ one can get blocks $3 \times 3$  and $1 \times 1$.

\subsection{[5,5,3,3,1,1]}
\begin{equation}
\mathcal{R}_{[5,5,3,3,1,1]} =
 {\rm diag}(
-q^{\varkappa_{[5,4,2,1]}}, \
 q^{\varkappa_{[5,4,2,1]}}, \
 q^{\varkappa_{[5,3,3,1]}}, \
 -q^{\varkappa_{[5,3,2,2]}}, \
 -q^{\varkappa_{[4,4,3,1]}})
= \\ {\rm diag}(
-q^{-6} , \
 q^{-6} , \
 q^{-4}, \
 -q^{-2}, \
 -q^{-2} )
\end{equation}
There is just one pair of coinciding eigenvalues ($\lambda_4 = \lambda_5$ accidentally). After a rotation with
\begin{equation}
\boxed{\begin{array}{ll}
	c_{45}^{[5,5,3,3,1,1]} = \frac{1}{\sqrt{2}}, & s_{45}^{[5,5,3,3,1,1]} = -\frac{1}{\sqrt{2}}
	\end{array}}
\end{equation}
we get
\begin{equation}
\mathcal{U'}_{[5,5,3,3,1,1]} =
\left(
\begin{array}{ccccc}
-\frac{q^2}{q^4+1} & 0 & 0 & \frac{\sqrt{q^4-q^2+1} \sqrt{q^4+q^2+1}}{q^4+1} & 0 \\
0 & \frac{q^2}{\left(q^2+1\right)^2} & -\frac{\sqrt{2} q \sqrt{q^4+q^2+1}}{\left(q^2+1\right)^2} & 0 & -\frac{q^4+q^2+1}{\left(q^2+1\right)^2} \\
0 & -\frac{\sqrt{2} q \sqrt{q^4+q^2+1}}{\left(q^2+1\right)^2} & \frac{q^4+1}{\left(q^2+1\right)^2} & 0 & -\frac{\sqrt{2} q
	\sqrt{q^4+q^2+1}}{\left(q^2+1\right)^2} \\
-\frac{\sqrt{q^4-q^2+1} \sqrt{q^4+q^2+1}}{q^4+1} & 0 & 0 & -\frac{q^2}{q^4+1} & 0 \\
0 & -\frac{q^4+q^2+1}{\left(q^2+1\right)^2} & -\frac{\sqrt{2} q \sqrt{q^4+q^2+1}}{\left(q^2+1\right)^2} & 0 & \frac{q^2}{\left(q^2+1\right)^2} \\
\end{array}
\right).
\end{equation}
One can see $3 \times 3 $ and $2 \times 2 $ blocks in this matrix.

\subsection{[5,4,4,2,2,1]}
\begin{multline}
\mathcal{R}_{[5,4,4,2,2,1]} =
 {\rm diag}(
-q^{\varkappa_{[5,4,2,1]}} , \
q^{\varkappa_{[5,4,2,1]}} , \
 -q^{\varkappa_{[5,3,2,2]}}, \
 -q^{\varkappa_{[4,4,3,1]}}, \
 q^{\varkappa_{[4,4,2,2]}} , \
 q^{\varkappa_{[4,4,2,2]}})
 = \\
 {\rm diag}(
 -q^{-6}, \
 q^{-6} , \
 -q^{-2} , \
 -q^{-2} , \
  1, \
 1).
\end{multline}
There are two pairs of coinciding eigenvalues: $\lambda_3 = \lambda_4$ (accidentally), $\lambda_5 = \lambda_6$.
\begin{equation}
\boxed{\begin{array}{ll}
	c_{34}^{[5,4,4,2,2,1]} = \frac{1}{\sqrt{2}}, & s_{34}^{[5,4,4,2,2,1]} = -\frac{1}{\sqrt{2}},
	\end{array}}
\end{equation}
\begin{equation}
\begin{array}{ll}
c_{1,56}^{[5,4,4,2,2,1]} = \sqrt{\frac{\left(q^4+q^2+1\right) \left(q^8+1\right)}{2(q^{12}+2 q^8-q^6+2 q^4+1)}}, &
s_{1,56}^{[5,4,4,2,2,1]} = -\frac{q^4+1}{\sqrt{2} \sqrt{q^8+q^6+2 q^4+\frac{1}{q^4-q^2+1}}} , \\
c_{2,56}^{[5,4,4,2,2,1]} = \frac{q^4+1}{\sqrt{2} \sqrt{q^8+q^6+2 q^4+\frac{1}{q^4-q^2+1}}}, &
s_{2,56}^{[5,4,4,2,2,1]}= \frac{\left(q^4+q^2+1\right) \left(q^8+1\right)}{\sqrt{2} \sqrt{\left(q^{12}+2 q^8-q^6+2 q^4+1\right) \left(q^{12}+q^{10}+q^8+q^4+q^2+1\right)}} .\\
\end{array}
\end{equation}

%
%
\begin{multline}
\mathcal{U'}_{[5,4,4,2,2,1]} = \\
\left(
\begin{array}{cccccc}
-\frac{q^4}{q^8+q^6+2 q^4+q^2+1} & 0 & x_{13} & 0 & 0 & x_{16} \\
0 & \frac{q^4}{q^8+q^6+q^2+1} & 0 &x_{24} & x_{25} & 0 \\
x_{13} & 0 & \frac{q^2}{q^4+1}-1 & 0 & 0 & \frac{q}{\sqrt{q^4+q^2+1}} \\
0 & -x_{24} & 0 & \frac{q^4+q^2+1}{\left(q^2+1\right)^2} & \frac{q \sqrt{\left(q^4-q^2+1\right) \left(q^8+1\right)}}{q^8+q^6+q^2+1} & 0 \\
0 & x_{25} & 0 & -\frac{q \sqrt{\frac{q^8+1}{q^4-q^2+1}}}{\left(q^2+1\right)^2} & -\frac{q^6+q^2}{q^8+q^6+q^2+1} & 0 \\
-x_{16} & 0 & -\frac{q}{\sqrt{q^4+q^2+1}} & 0 & 0 & \frac{q^2}{q^4+q^2+1} \\
\end{array}
\right),
\end{multline}
where
\begin{equation}
\begin{array}{ll}
x_{13} = \frac{q \sqrt{\left(q^2+1\right) \left(q^8+q^6+2 q^4+q^2+2\right) q^2+1}}{q^8+q^6+2 q^4+q^2+1}, &
x_{16} = \frac{\sqrt{q^8+q^6+q^4+q^2+1}}{q^4+q^2+1}, \\
x_{24} = \frac{q \sqrt{q^{12}+q^8+q^6+q^4+1}}{q^8+q^6+q^2+1}, &
x_{25} = -\frac{\sqrt{\left(q^4-q^2+1\right) \left(q^8+1\right) \left(q^{12}+q^8+q^6+q^4+1\right)}}{\left(q^6+1\right)^2}.
\end{array}
\end{equation}
One can see that $\mathcal{U'}_{[5,4,4,2,2,1]} $ has two blocks $3 \times 3$.
\subsection{[11,6,1]}
\begin{multline}
\mathcal{R}_{[11,6,1]} =
 {\rm diag}(
q^{\varkappa_{[8,4]}} , \
 q^{\varkappa_{[8,4]}}, \
 -q^{\varkappa_{[8,3,1]}}, \
 -q^{\varkappa_{[7,5]}} , \
 -q^{\varkappa_{[7,4,1]}} , \
 q^{\varkappa_{[7,4,1]}} )
= \\
 {\rm diag}(
q^{-30} , \
 q^{-30}, \
 -q^{-26} , \
 -q^{-26} , \
-q^{-21}, \
q^{-21}).
\end{multline}
There are two pairs of coinciding eigenvalues: $\lambda_1 = \lambda_2$ and $\lambda_3 = \lambda_4$ (accidentally).
%
%
%
\begin{equation}
\begin{array}{l}
c_{1,12}^{[11,6,1]} =\sqrt{\frac{\left(q^2+q+1\right) \left(q^4+q^3+q^2+q+1\right) \left(q^6+q^3+1\right) \left((q-1) q ((q-1) q+1) \left(q^2+1\right) \left(q^2+q+1\right)
		\left(q^4-q^2+1\right)+1\right)}{2(\left(q^2+1\right) \left(q^4+1\right) \left(q^8-q^4+1\right) \left(q^8+q^6+2 q^4+q^2+2\right) q^2+1)}},\\
s_{1,12}^{[11,6,1]} = \sqrt{\frac{((q-1) q+1) \left(q^6-q^3+1\right) \left((q-1) q \left(q^2+1\right)+1\right) \left(q (q+1) ((q-1) q+1) \left(q^2+1\right) \left(q^2+q+1\right)
		\left(q^4-q^2+1\right)+1\right)}{2(\left(q^2+1\right) \left(q^4+1\right) \left(q^8-q^4+1\right) \left(q^8+q^6+2 q^4+q^2+2\right) q^2+1)}},\\
c_{2,12}^{[11,6,1]} = \sqrt{\frac{((q-1) q+1) \left(q^6-q^3+1\right) \left((q-1) q \left(q^2+1\right)+1\right) \left(q (q+1) ((q-1) q+1) \left(q^2+1\right) \left(q^2+q+1\right)
		\left(q^4-q^2+1\right)+1\right)}{2(\left(q^2+1\right) \left(q^4+1\right) \left(q^8-q^4+1\right) \left(q^8+q^6+2 q^4+q^2+2\right) q^2+1)}} ,\\
s_{2,12}^{[11,6,1]} = \frac{\sqrt{\left(q^2+q+1\right) \left(q^6+q^3+1\right)} \left(\left(q^2+1\right) \left(q^4+1\right) \left(q^8-q^3+1\right) q^2+1\right)}{\sqrt{2}
	\sqrt{\left(q^4+q^3+q^2+q+1\right) \left((q-1) q ((q-1) q+1) \left(q^2+1\right) \left(q^2+q+1\right) \left(q^4-q^2+1\right)+1\right)
		\left(\left(q^2+1\right) \left(q^4+1\right) \left(q^8-q^4+1\right) \left(q^8+q^6+2 q^4+q^2+2\right) q^2+1\right)}}.\\
\end{array}
\end{equation}
and
\begin{equation}
\boxed{\begin{array}{ll}
	c_{34}^{[11,6,1]} = \frac{1}{\sqrt{2}}, & s_{34}^{[11,6,1]} = -\frac{1}{\sqrt{2}}.
	\end{array}}
\end{equation}
After the rotations we'll get a matrix
\begin{equation}
\mathcal{U'}_{[11,6,1]} =
\left(
\begin{array}{cccccc}
x_{11} & 0 & x_{13} & 0 & x_{15} & 0 \\
0 &x_{22} & 0 &x_{24} & 0 & x_{26} \\
x_{13} & 0 & x_{33} & 0 &x_{35} & 0 \\
0 & -x_{24} & 0 & x_{44} & 0 & -x_{46}\\
x_{15} & 0 & x_{35} & 0 &x_{55} & 0 \\
0 & x_{26} & 0 & x_{46} & 0 & x_{66} \\
\end{array}
\right),
\end{equation}
%
%
where
\begin{equation}
\begin{array}{ll}
x_{11} = \frac{q^4 \left(q^4+q^3+q^2+q+1\right)}{\left(q^2+q+1\right) \left(q^4+1\right) \left(q^6+q^3+1\right)}, &
x_{13} = -\frac{q^2 \sqrt{\frac{\left(q^2+1\right) \left(q^4+1\right) \left(q^8+q^3+1\right) q^2+1}{\left(q^2+q+1\right)
			\left(q^6+q^3+1\right)}}}{\left(q^4+1\right) \left((q-1) q \left(q^2+1\right)+1\right)}, \\
x_{15} = \frac{\sqrt{\left(q^{16}+q^{14}+2 q^{12}+2 q^{10}+2 q^8+2 q^6+2 q^4+q^2+1\right) \left(\left(q^2+1\right) \left(q^4+1\right) \left(q^8+q^3+1\right)
		q^2+1\right)}}{\left(q^2+q+1\right) \left(q^4+1\right) \left(q^6+q^3+1\right) \left((q-1) q \left(q^2+1\right)+1\right)}, &
x_{22} = \frac{q^4 \left((q-1) q \left(q^2+1\right)+1\right)}{((q-1) q+1) \left(q^4+1\right) \left(q^6-q^3+1\right)}, \\
x_{24} = \frac{q^2 \sqrt{\frac{\left(q^2+1\right) \left(q^4+1\right) \left(q^8-q^3+1\right) q^2+1}{((q-1) q+1) \left(q^6-q^3+1\right)}}}{\left(q^4+1\right)
	\left(q^4+q^3+q^2+q+1\right)}, &
x_{33} = \frac{((q-1) q+1) \left(q^6-q^3+1\right)}{\left(q^4+1\right) \left((q-1) q \left(q^2+1\right)+1\right)}, \\
x_{35} = \frac{q^2}{\sqrt{\frac{\left(q^2+q+1\right) \left(q^4+1\right) \left(q^6+q^3+1\right)}{q^{12}+q^{10}+q^8+q^6+q^4+q^2+1}} \left((q-1) q
	\left(q^2+1\right)+1\right)}	, &
x_{44} = -\frac{\left(q^2+q+1\right) \left(q^6+q^3+1\right)}{\left(q^4+1\right) \left(q^4+q^3+q^2+q+1\right)}, \\
x_{46} = \frac{q^2}{\left(q^4+q^3+q^2+q+1\right) \sqrt{\frac{((q-1) q+1) \left(q^4+1\right) \left(q^6-q^3+1\right)}{q^{12}+q^{10}+q^8+q^6+q^4+q^2+1}}}, &
x_{55} = \frac{q^7+q^5}{\left(q^2+1\right) \left(q^8+q^4+q^3+1\right) q^2+1}, \\
x_{66} = -\frac{q^7+q^5}{q^{12}+q^{10}+q^8-q^7+q^6-q^5+q^4+q^2+1}.\\
\end{array}
\end{equation}
$\mathcal{U'}_{[11,6,1]}$ contains two blocks $3 \times 3$.

\section{Conclusion \label{conclusion}}
We investigated all Racah matrices up to size $6 \times 6$ and some matrices of size $8 \times 8$ and $9 \times 9$ from cubes of representations $[2,1], \ [3,1], \  [3,2], \  [4,1]$  and $ [4,2]$. We found out that some of them could be transformed into block-diagonal ones. The essential condition was that the corresponding $\mathcal{R}$-matrix contained a pair of accidentally coinciding eigenvalues. The additional condition emerged for matrices of size $6 \times 6$ and larger. In this case corresponding $\mathcal{R}$-matrices should also contain repetitive eigenvalues: one pair for $6 \times 6$ matrices and two pairs for $8 \times 8$. We couldn't make such conclusion for $9 \times 9$ matrices because we had both successful and unsuccessful examples in the case of two pairs of repetitive eigenvalues and one pair of accidentally coinciding eigenvalues.

The transformations that made the Racah matrices (and therefore $\mathcal{R}_2$-matrices) block-diagonal were rotations in the sectors of coinciding eigenvalues (accidental and repetitive). Particular examples showed that the angle of rotation in "accidental" sector was always $\pm \frac{\pi}{4}$. Since such rotation mixed vectors from different irreducible representations, the resulting matrix didn't satisfy the definition (\ref{racahdef}) and no longer was a Racah matrix, but the $\mathcal{R}_2$-matrix it produced was suitable for calculations.

The practical impact of this work is that it gives us hope to use the eigenvalue hypothesis \cite{eivcon} to calculate Racah matrices even in the case of coinciding eigenvalues. It's possible because the resulting blocks in the corresponding $\mathcal{R}$-matrix don't contain coinciding eigenvalues. It's still a question how to determine which eigenvalues go to each particular block.

In case of large matrices (of size $9 \times 9$ and larger) more complex cases emerged: when more than a pair of eigenvalues coincided. Such groups of coinciding eigenvalues could consist of both accidental and repeating eigenvalues. The problem was that there was more than one angle of rotation in this case and it's a lot more difficult to determine  the right rotation matrix.

\section{Acknowledgements}
We would like to thank A.Morozov and A.Mironov for useful discussions. This work was funded by the Russian Science Foundation (Grant No. 16-12-10344).

\end{document}